\documentclass{article}
\usepackage{arxiv}
\usepackage{iftex}
\ifPDFTeX\usepackage[utf8]{inputenc}\fi
\usepackage[T1]{fontenc}
\usepackage{amsmath,amssymb}
\usepackage{graphicx}
\usepackage{booktabs,array}
\usepackage[round,authoryear]{natbib}
\usepackage{xcolor}
\usepackage{url}
\usepackage{hyperref}
\usepackage{caption}
\usepackage[section]{placeins}
\usepackage{microtype}
\renewcommand{\headeright}{}
\renewcommand{\undertitle}{}
\renewcommand{\shorttitle}{\textit{Radfar et al. (2026)} --- AutoCF}
\date{}
\hypersetup{pdfauthor={Soheil Radfar, Faezeh Maghsoodifar, Ning Lin, Hamed Moftakhari},
  pdftitle={AutoCF: An Automated LLM-Assisted Ecosystem for Compound Flood Simulation, Evaluation, and Impact Attribution},colorlinks=true,linkcolor=blue,citecolor=blue,urlcolor=blue}
\title{AutoCF: An Automated LLM-Assisted Ecosystem for Compound Flood Simulation, Evaluation, and Impact Attribution}
\author{
\href{https://orcid.org/0000-0003-0177-9733}{\includegraphics[scale=0.06]{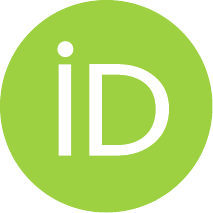}\hspace{1mm}Soheil Radfar*}\\
Department of Civil and Environmental Engineering\\
Princeton University, NJ, USA\\
\texttt{sradfar@princeton.edu}\\
\And
\href{https://orcid.org/0000-0002-6522-0712}{\includegraphics[scale=0.06]{orcid.pdf}\hspace{1mm}Faezeh Maghsoodifar}\\
Department of Civil, Construction\\
and Environmental Engineering\\
The University of Alabama, AL, USA\\
\texttt{fmaghsoodifar@crimson.ua.edu}\\
\AND
\href{https://orcid.org/0000-0002-5571-1606}{\includegraphics[scale=0.06]{orcid.pdf}\hspace{1mm}Ning Lin*}\\
Department of Civil and Environmental Engineering\\
Princeton University, NJ, USA\\
\texttt{nlin@princeton.edu}\\
\And
\href{https://orcid.org/0000-0003-3170-8653}{\includegraphics[scale=0.06]{orcid.pdf}\hspace{1mm}Hamed Moftakhari}\\
Department of Civil, Construction\\
and Environmental Engineering\\
The University of Alabama, AL, USA\\
\texttt{hmoftakhari@eng.ua.edu}\\
}
\begin{document}
\raggedbottom
\maketitle
\begin{abstract}
Compound coastal flooding (CCF) arises from interacting coastal, precipitation, and river processes, yet modeling workflows often separate simulation, evaluation, and impact analysis. We present AutoCF, an automated ecosystem integrating data harmonization, model construction, observational evaluation, exposure analysis, complete factorial driver attribution, and cross-platform execution. The automated Hurricane Harvey simulation achieves a median root mean square error of 0.147 m and correlation of 0.951 across eight observational gauges, and a correlation of 0.942 with 55 high-water marks. Maximum water-level fields from CPU and GPU implementations agree within 0.01 m for 98.8\% of cells. Attribution analysis shows that during Harvey, precipitation dominated building and population exposure, whereas coastal forcing becomes increasingly important for deep and persistent inundation. The introduced Driver Impact Shift metric further quantifies whether each driver contributed disproportionately to societal consequences relative to its flooded-area contribution. Overall, AutoCF provides a reproducible pathway from CCF model construction to evaluation and driver-specific impact interpretation.
\end{abstract}
\keywords{Compound coastal flooding; automated flood modeling; hydrodynamic simulation; impact attribution; reproducible scientific workflows}

\section{Introduction}
\label{sec:1}

Flooding in coastal and deltaic regions can result from the concurrent or sequential effects of heavy precipitation, river discharge, and elevated coastal water levels \citep{radfar2024}. Their interaction can generate compound coastal flooding (CCF) with inundation patterns and consequences that differ substantially from those associated with individual drivers \citep{moftakhari2017,wahl2015}. Tropical and extratropical storms are particularly important because they can simultaneously produce heavy rainfall, enhanced river flow, strong winds, and elevated coastal water levels over overlapping spatial and temporal scales \citep{grimley2026,lai2021}. Consequently, CCF assessment increasingly relies on integrated hydrodynamic modeling that represents interactions among pluvial, fluvial, and coastal processes rather than treating each source independently \citep{bates2021,jafarzadegan2021,radfar2026}. Recent advances in global terrain, hydrography, meteorological, hydrological, and oceanographic datasets have also expanded flood modeling from individual catchments toward continental and global applications \citep{nazari2026,wing2024}. However, flood impacts remain strongly controlled by local terrain, river geometry, surface properties, boundary conditions, and the spatial interaction among flood drivers, making both model resolution and observational evaluation important for credible applications \citep{grimley2025,maghsoodifar2025,sebastian2021}.

Increasing model capability and data availability have not eliminated the practical complexity of constructing and analyzing CCF simulations. Applications still require heterogeneous datasets to be discovered, harmonized, spatially processed, assigned consistent units and vertical references, translated into model inputs, executed, and evaluated. Flood-specific software has progressively automated parts of this process. LFPtools streamlines preparation of inundation models from heterogeneous datasets \citep{sosa2020}, Delft Dashboard supports rapid construction of hydrodynamic model schematizations \citep{vanormondt2020}, HydroMT provides configuration-driven and reproducible model building and analysis \citep{eilander2023a}, and FIMserv integrates flood-inundation generation with hydrological data acquisition and observational evaluation \citep{baruah2025}. A globally applicable CCF framework has further demonstrated that local SFINCS models can be constructed automatically from global datasets while accounting for interactions among pluvial, fluvial, and coastal drivers \citep{eilander2023b}. Yet model construction is only one component of a complete CCF analysis. Observational evaluation is needed to determine whether the assembled terrain, forcing, and boundary conditions reproduce the event, while exposure and driver attribution are needed to determine where consequences occur and which processes contribute to them. For example, \citep{grimley2025} showed that during Hurricane Florence, compound processes accounted for only 15.3\% of the total flooded area but 31\% of building exposure, while coastal processes contributed 5.9\% of flooded area but 14\% of building exposure. This demonstrates that attribution based on flood extent alone may not reflect the drivers of societal exposure and motivates a workflow that connects model preparation, simulation, evaluation, and impact attribution within a common reproducible environment.

Here, we introduce AutoCF, an automated, large language model (LLM)-assisted ecosystem for CCF simulation, impact attribution, and evaluation. AutoCF supports an end-to-end workflow that acquires and harmonizes terrain, land-cover, soil, atmospheric, coastal, river, and observational data; records source and vertical-datum provenance; constructs and executes CCF simulations; evaluates modeled water levels and inundation against observations; and quantifies exposure and driver-specific impacts. The existing implementation uses HydroMT-SFINCS \citep{eilander2023a} for hydraulic model construction and the Super-Fast INundation of CoastS model (SFINCS; \citep{leijnse2021}) for hydrodynamic simulation. AutoCF extends this modeling chain through automated evaluation, exposure analysis, exact factorial attribution of coastal, precipitation, and river forcing, flood-duration analysis, and reproducible execution across computing environments.

The desktop distribution includes a local LLM-assisted interface that provides workflow guidance and interpretation of completed analyses. The portable high-performance computing (HPC) distribution retains the scientific model-building and post-processing workflow. Together, these components provide a reproducible pathway from environmental data and CCF simulation to observational evaluation and driver-specific impact interpretation across supported desktop and HPC environments. In its current implementation, AutoCF is designed primarily for event-based hindcasting, where historical precipitation, coastal water levels, river discharge, and other available observations or reanalysis products are assembled to reconstruct CCF inundation and its impacts for a specified event and study domain.

\section{AutoCF framework}
\label{sec:2}

\begin{figure}[!htbp]
\centering
\includegraphics[width=\linewidth,height=0.76\textheight,keepaspectratio]{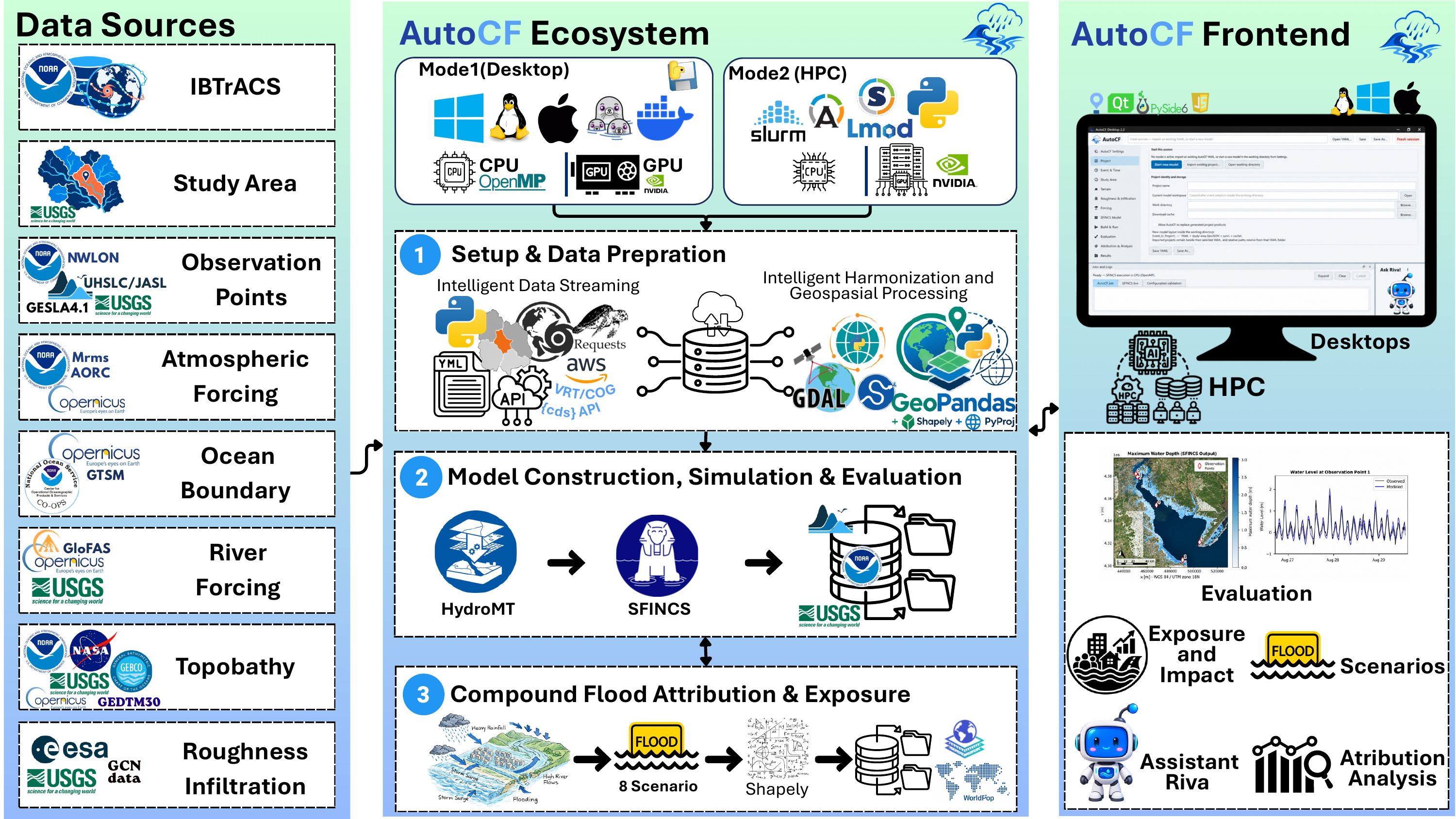}
\caption{Overview of the AutoCF architecture and workflow. Environmental datasets are acquired and harmonized for project setup and data preparation, followed by hydraulic model construction, simulation, and evaluation. Post-processing modules perform CCF attribution and exposure analysis. The desktop frontend provides access to model setup, results, and LLM-assisted interpretation, whereas the portable HPC distribution exposes the scientific workflow through command-line operations and excludes the desktop and local language-model components.}
\label{fig:1}
\end{figure}

AutoCF (Figure~\ref{fig:1}) is a configuration-driven environment for end-to-end CCF modeling and analysis. Starting from a user-defined study domain and simulation period, it coordinates the acquisition and harmonization of terrain, surface, atmospheric, coastal, river, and observational data; hydraulic model construction and execution; observational evaluation; and impact and driver-attribution analyses. Scientific settings, data sources, and processing choices are retained in a human-readable project configuration, allowing the workflow to be reproduced independently of the graphical interface. AutoCF performs the data acquisition, harmonization, geospatial preprocessing, forcing preparation, and provenance tracking, then uses HydroMT-SFINCS \citep{eilander2023a} to assemble the hydraulic model from these prepared inputs. SFINCS \citep{leijnse2021} performs the hydrodynamic simulation, after which AutoCF conducts the evaluation, impact analysis, and attribution.

The workflow is modular, allowing major input components to be prepared and inspected before model construction and reused in subsequent stages. The same project configuration controls simulation, evaluation, exposure analysis, and CCF attribution, maintaining consistency across the analysis chain. Existing SFINCS models can also be imported directly for evaluation and post-processing without repeating the complete model-construction workflow. Hydraulic parameterization remains user-controlled. AutoCF provides reproducible defaults, consistency checks, input provenance, and explicit configuration options. This structure combines automation with the flexibility needed to incorporate local data and modeling decisions.

\subsection{System architecture and design principles}
\label{sec:2.1}

AutoCF is organized into five interacting logical layers: the desktop presentation layer, canonical configuration layer, scientific orchestration layer, post-processing and interpretation layer, and runtime and deployment layer. The desktop application supports project management, spatial interaction, configuration, data preparation, model execution, validation, and access to scientific results. User-defined scientific settings are passed through the canonical project configuration, keeping the modeling workflow independent of transient interface states and allowing projects to be inspected, archived, modified, and reused across sessions and computing environments.

The canonical configuration provides the common interface between the user application and the scientific workflow. It centrally manages project-relative paths, defaults, source metadata, unit conversions, and compatibility rules. During project development, configurations can be assembled progressively as data sources and model settings are defined. Before model construction and execution, stricter dependency and consistency checks are applied to ensure that the resolved configuration is scientifically and technically complete. This separation supports interactive project development while preserving rigorous validation of the final model configuration.

The scientific orchestration layer converts the resolved configuration into the datasets, catalogs, and model-building instructions required for hydraulic simulation. Preparation stages for terrain, surface properties, and atmospheric forcing can be executed independently, allowing intermediate products to be inspected before model construction. The orchestration layer also validates required inputs and the resulting model configuration before execution. Post-processing modules operate directly on saved simulation products to perform observational evaluation, exposure analysis, and CCF attribution. Scientific results are then organized separately from intermediate and technical files so that the interface presents the principal figures, tables, and reports required for interpretation.

AutoCF also separates the native desktop interface from the controlled scientific runtime. Geospatial processing, model construction, and hydraulic simulation are executed within a managed environment with defined software dependencies, while the desktop application remains responsible for user interaction and workflow control. This architecture supports desktop, containerized, and HPC deployment using a common project configuration and scientific workflow. Platform-specific numerical agreement and computational performance are verified explicitly in the benchmark experiments described in Section~\ref{sec:3.5}.

These design choices allow AutoCF projects to progress modularly from data preparation to simulation and post-processing while preserving configuration, provenance, and validated intermediate products. Computationally expensive acquisitions and completed workflow stages can be reused when their defining inputs remain unchanged, with detailed caching, resumability, and run-state management described in Sections~\ref{sec:2.4} and \ref{sec:2.7}.

\subsection{Automated data acquisition and harmonization}
\label{sec:2.2}

\subsubsection{Event and study-area definition}
\label{sec:2.2.1}

AutoCF supports both U.S.-specific and globally applicable data pathways. A project can be initialized for a user-defined simulation period in Coordinated Universal Time (UTC) or from a tropical cyclone selected through the National Oceanic and Atmospheric Administration (NOAA) International Best Track Archive for Climate Stewardship (IBTrACS; \citep{knapp2010}). The study domain can be drawn interactively or imported as a polygon. Within the same spatial framework, users can define coastal boundaries, river inflow locations, observation points, high-water marks (HWMs), and spatial roughness overrides, ensuring that these features retain consistent roles throughout subsequent processing and simulation.

For U.S. applications, watershed boundaries can be obtained from the United States Geological Survey (USGS) Watershed Boundary Dataset (WBD; \citep{jones2022}), including nested drainage units identified by hydrologic unit codes (HUCs). River and water-level stations can also be discovered from federal monitoring networks. Where supported, spatial searches are constrained to the actual model polygon rather than the enclosing bounding box, reducing the inclusion of stations that are geographically nearby but outside the modeled hydrologic domain.

Acquired datasets are passed through source-specific harmonization before entering the CCF workflow. Processing can include spatial and temporal subsetting, coordinate transformation, variable standardization, unit conversion, missing-value treatment, physical-range checks, and conversion to the format required for model construction. Provider information and processing metadata are retained with the prepared products, while compatible cached data can be reused for repeated requests with unchanged spatial, temporal, and processing specifications.

\subsubsection{Land-surface properties and infiltration}
\label{sec:2.2.2}

AutoCF derives spatially variable surface roughness from the European Space Agency (ESA) WorldCover 10 m 2021 product \citep{zanaga2022}, the U.S. Annual National Land Cover Database (Annual NLCD; \cite{dewitz2021}), or a user-supplied land-cover raster. Land-cover classes are converted to Manning roughness coefficients using configurable mappings, with optional polygon-based overrides where local information is available. AutoCF acquires or clips only the source data needed for the study domain, mosaics tiled products where necessary, and transfers the prepared roughness data to the hydraulic representation during model construction.

Multiple infiltration pathways are supported to accommodate differences in data availability and application requirements. Simulations can use the global GCN250 Curve Number dataset \citep{jaafar2019}, a user-defined Curve Number raster, or a higher-detail U.S. pathway combining Annual NLCD with the U.S. Department of Agriculture Natural Resources Conservation Service Soil Survey Geographic Database (SSURGO). GCN250 provides globally consistent Soil Conservation Service Curve Number (SCS-CN) values at 250 m resolution, while the SSURGO pathway combines land cover, hydrologic soil groups, and hydraulic soil properties to derive Curve Number, storage, and recovery characteristics.

For SSURGO-based infiltration, antecedent moisture conditions can be represented as dry, average, or wet. Dual hydrologic soil groups and missing classifications follow configurable rules, while saturated hydraulic conductivity is aggregated over the selected soil depth using thickness- and component-weighted calculations. The resulting infiltration fields are then transferred to the model representation through HydroMT-SFINCS.

\subsubsection{Atmospheric forcing}
\label{sec:2.2.3}

AutoCF supports multiple atmospheric forcing pathways for precipitation, wind, and surface pressure. In the CCF workflow, precipitation provides direct pluvial forcing, while wind and surface pressure represent local atmospheric forcing within the hydraulic model domain. These atmospheric fields are not required to generate the regional storm surge signal when tide and surge are prescribed separately through the coastal water level boundary described in Section~\ref{sec:2.2.4}.

For U.S. applications, the NOAA Analysis of Record for Calibration (AORC; \citep{fall2023}) provides spatially distributed precipitation, wind, and pressure, while the European Centre for Medium-Range Weather Forecasts Reanalysis version 5 (ERA5; \citep{hersbach2020}) provides globally available forcing. Precipitation can also be supplied from NOAA Multi-Radar Multi-Sensor (MRMS) Quantitative Precipitation Estimation files \citep{zhang2016}. User-supplied precipitation, wind, and pressure fields can be substituted when local or externally prepared datasets are preferred.

Source-specific preprocessing is applied before atmospheric data enter the hydraulic model. AORC is spatially and temporally subset from its cloud-hosted archive, with explicit treatment of its end-labelled hourly precipitation accumulation, followed by variable and unit standardization, physical-range checks, missing-cell handling, and spatial-coverage validation. ERA5 is acquired through user-managed Copernicus access and undergoes equivalent temporal, spatial, unit, and plausibility checks before model construction. These procedures ensure that differences among atmospheric data providers are resolved before forcing fields are passed to the CCF simulation.

It should be noted that since gridded atmospheric products may not fully resolve the compact wind and pressure structure of intense TCs, particularly when coarse reanalysis products are used, higher-resolution or TC-specific atmospheric forcing is recommended when local wind setup and pressure-driven water-level responses are important to the application.

\subsubsection{Coastal and river boundary forcing}
\label{sec:2.2.4}

AutoCF supports both observational and model-based sources for coastal water-level boundaries. For U.S. applications, water levels can be obtained from NOAA Center for Operational Oceanographic Products and Services (CO-OPS) stations. Global pathways include the Global Extreme Sea Level Analysis (GESLA-4.1; \citep{haigh2023}), the University of Hawai'i Sea Level Center (UHSLC) and Joint Archive for Sea Level (JASL; \citep{caldwell2015}), and the Global Tide and Surge Model (GTSM; \citep{muis2020}). Users can also provide their own coastal water-level time series and boundary locations.

River discharge can be acquired from the USGS National Water Information System (NWIS) for U.S. applications or from the Global Flood Awareness System (GloFAS; \citep{harrigan2020}) for broader geographic coverage. User-defined discharge series and inflow locations are also supported. Gauge discovery is constrained to the model domain, and acquired discharge records are standardized to consistent International System of Units (SI) before model construction.

\subsubsection{Observation data, quality control, and provenance}
\label{sec:2.2.5}

Observation records used for model evaluation are stored with their provider metadata and available vertical-reference information and are passed to the evaluation workflow described in Section~\ref{sec:2.5}. Across all prepared datasets, AutoCF retains source and processing provenance and applies quality-control checks for spatial and temporal coverage, units, missing values, datum compatibility, and physical plausibility. Key intermediate products can also be inspected before model construction so that data issues can be identified before they propagate into the hydraulic simulation.

\subsection{Terrain and vertical-datum management}
\label{sec:2.3}

Terrain is handled separately because CCF simulations are highly sensitive to elevation errors, discontinuities, and inconsistencies in vertical reference. AutoCF uses a hierarchical terrain workflow with a required primary digital elevation model (DEM) and optional additional sources for filling areas without valid elevation. The project configuration records the terrain source, native vertical datum, processing resolution, buffer, and datum-handling method so that the final model surface remains traceable to its inputs.

Built-in U.S. terrain sources include the NOAA Continuously Updated Digital Elevation Model (CUDEM) and the USGS 3D Elevation Program (3DEP). Global options include the General Bathymetric Chart of the Oceans \citep{gebcobathymetriccompilationgroup2026}, the Global Ensemble Digital Terrain Model at 30 m resolution (GEDTM30; \citep{ho2025}), and the Copernicus Digital Elevation Model GLO-30 (Copernicus DEM GLO-30; \citep{opentopography2021}). User-provided terrain rasters can replace these products where higher-resolution or locally validated data are available.

The primary terrain source defines the target vertical reference. When the optional gap-fill source uses a different datum, automatic mode first attempts a rigorous grid-based conversion using a global vertical datum transformations, Transformez \citep{continuousdem2026}. Otherwise, a local vertical offset can be estimated from overlapping valid elevations, with the offset and supporting diagnostics retained in the project metadata. Users can also require a rigorous transformation or provide a known offset. This hierarchy enables gap filling while preserving explicit control over vertical-reference uncertainty.

Terrain datum information is propagated to observational evaluation so that modeled and observed water levels are compared using compatible vertical references. Absolute-datum statistics are reported only when the modeled and observed elevations use compatible datums or an explicit, documented offset; otherwise, the comparison is identified as relative alignment. AutoCF also records temporal compatibility warnings when available terrain metadata indicate that the elevation source may postdate the simulated event, preserving this limitation in the model provenance.

\subsection{Automated model construction and execution}
\label{sec:2.4}

Once terrain, surface properties, and forcing datasets have been prepared, AutoCF converts the resolved project configuration into a complete hydraulic model. Model construction and simulation are maintained as separate stages, allowing the generated model to be inspected before execution and permitting previously constructed models to be rerun without repeating data acquisition or preprocessing.

\subsubsection{Hydraulic model configuration}
\label{sec:2.4.1}

AutoCF supports both regular-grid and SFINCS subgrid configurations. Users define the computational resolution and may allow AutoCF to select an appropriate Universal Transverse Mercator (UTM) projection or provide a coordinate reference system (CRS). The project configuration also specifies the simulation period, numerical controls, output frequency, and variables retained for subsequent evaluation and analysis.

For regular-grid simulations, elevation, roughness, infiltration, and forcing are represented directly on the computational grid. In subgrid mode, higher-resolution terrain and roughness information are incorporated into precomputed lookup tables while the hydrodynamic equations are solved on a coarser grid. This retains information on unresolved topographic and friction variability while reducing computational cost relative to a uniformly fine hydraulic grid \citep{vanormondt2025}. AutoCF exposes the sampling controls required for subgrid construction while preserving the terrain provenance described in Section~\ref{sec:2.3}.

Hydraulic and output parameters are controlled through the project configuration, including numerical coefficients, thresholds, infiltration and viscosity settings, and retained diagnostic variables. AutoCF checks parameter dependencies and compatibility before model construction and execution, and the generated configuration remains available for inspection and reproducibility.

\subsubsection{Automated model assembly}
\label{sec:2.4.2}

AutoCF generates a HydroMT data catalog and build configuration from the datasets prepared through its acquisition, harmonization, and geospatial-processing stages and then invokes HydroMT-SFINCS to assemble the SFINCS model. Using these AutoCF-prepared products and build instructions, HydroMT-SFINCS creates the computational grid and masks, terrain representation, roughness and infiltration fields, coastal boundaries, river inflows, spatial atmospheric forcing, and observation locations. Subgrid lookup tables are generated when the subgrid formulation is selected. This automated construction follows the reproducible model-building approach previously demonstrated for globally applicable CCF simulations \citep{eilander2023b}.

Model assembly uses the harmonized products generated in the preceding preparation stages, preserving their terrain hierarchy, vertical-reference information, forcing metadata, and source provenance. Individual components can therefore be updated without reacquiring unrelated datasets. For example, changes to roughness or infiltration settings require the hydraulic model to be reconstructed while previously validated terrain and atmospheric products can be retained.

Preparation and construction stages can also be executed independently. This modular structure allows terrain, surface properties, atmospheric forcing, and the generated hydraulic model to be inspected before computationally expensive simulations are initiated.

\subsubsection{Pre-run validation and model execution}
\label{sec:2.4.3}

Before simulation, AutoCF verifies the constructed model and its forcing inputs. Required files are checked for availability and consistency, forcing coverage is compared with the configured simulation period, and dependencies among selected hydraulic options are evaluated. These checks establish that a successfully constructed model is also complete and internally consistent for execution.

Simulations are executed through the controlled scientific runtime while the desktop application remains responsive for monitoring and diagnostic feedback. Run logs retain standard output and error messages, and computational resource settings are passed explicitly to the runtime, including Open Multi-Processing (OpenMP) controls where supported. Hydraulic parameterization remains user-controlled, with selected settings and subsequent changes retained in the project configuration.

\subsubsection{Caching, resumability, and selective invalidation}
\label{sec:2.4.4}

AutoCF records completion states and input signatures for major preparation and model-building stages. These signatures track dependencies such as study-area geometry, terrain sources, surface properties, forcing configuration, and relevant input files. When a project setting changes, only the stages that depend on that setting are invalidated. For example, modifying the model domain requires spatially dependent products to be regenerated, whereas changing a roughness mapping retains valid terrain and forcing data while triggering reconstruction of the hydraulic model.

Previously validated products are reused when their signatures remain compatible with the current configuration, allowing interrupted preparation or construction to resume without repeating valid remote acquisitions. Cached downloads are accepted only after product-specific checks of file integrity, expected variables, spatial and temporal coverage, and physical ranges. Remote requests additionally use bounded retries and timeouts. This stage-aware caching and invalidation strategy reduces unnecessary computation and data transfer while preserving reproducibility during iterative model development.

\subsection{Model evaluation}
\label{sec:2.5}

AutoCF evaluates completed CCF simulations using spatial flood-depth products and point observations. The same evaluation workflow can be applied to models constructed within AutoCF or to imported SFINCS projects. The current implementation includes maximum flood-depth mapping, water-level time-series evaluation, and HWM comparison, with explicit treatment of observation coverage and vertical-datum compatibility.

\subsubsection{Maximum flood-depth calculation}
\label{sec:2.5.1}

For regular SFINCS grids, maximum flood depth is obtained directly from model output when available or calculated from the difference between maximum water-surface elevation and terrain elevation using

\begin{equation}
h_{\max}\left( x,y \right) = z_{\max}\left( x,y \right) - z_{b}\left( x,y \right)
\label{eq:1}
\end{equation}

where \(z_{\max}\) is the maximum simulated water-surface elevation and \(z_{b}\) is the terrain elevation. Cells below a configurable visualization threshold are masked from displayed flood maps while the underlying numerical output is retained. A configurable depth threshold can be applied to the derived visualization products, while the original SFINCS numerical output remains unchanged.

For subgrid simulations, the maximum simulated water surface is transferred to the higher-resolution subgrid terrain and flood depth is calculated relative to that terrain. Processing is performed blockwise to limit memory demand for large high-resolution domains, and the resulting depth field is retained as a georeferenced raster for subsequent evaluation and impact analysis.

\subsubsection{Water level time-series evaluation}
\label{sec:2.5.2}

Water-level observations can be obtained from the sources described in Section~\ref{sec:2.2.4} or supplied by the user. Candidate stations are screened against the study domain and simulation period before comparison with modeled water levels.

Vertical-reference handling follows the terrain and datum framework described in Section~\ref{sec:2.3}. Where modeled and observed elevations share a compatible absolute datum, direct water-level errors are calculated. Where only relative alignment is supported, AutoCF estimates a fixed median offset from an initial alignment period and applies it to the remaining record. The alignment period is excluded from validation statistics by default, and anomaly-based metrics are also reported relative to the established baseline. This preserves information on temporal model skill without presenting relative alignment as absolute elevation accuracy.

For \(n\) matched modeled water levels \(M_{i}\) and observations \(O_{i}\), the root mean square error (RMSE), mean absolute error (MAE), and the Pearson correlation coefficient \(r\) are defined as

\begin{equation}
RMSE = \sqrt{\frac{1}{n}\overset{n}{\underset{i = 1}{\sum_{}^{}}}\mspace{2mu}\left( M_{i} - O_{i} \right)^{2}}
\label{eq:2}
\end{equation}

\begin{equation}
MAE = \frac{1}{n}\overset{n}{\underset{i = 1}{\sum_{}^{}}}\mspace{2mu}\left| M_{i} - O_{i} \right|
\label{eq:3}
\end{equation}

\begin{equation}
\text{~Bias~} = \frac{1}{n}\overset{n}{\underset{i = 1}{\sum_{}^{}}}\mspace{2mu}\left( M_{i} - O_{i} \right)
\label{eq:4}
\end{equation}

and

\begin{equation}
r = \frac{\overset{n}{\underset{i = 1}{\sum_{}^{}}}\mspace{2mu}\left( M_{i} - \overline{M} \right)\left( O_{i} - \overline{O} \right)}{\sqrt{\overset{n}{\underset{i = 1}{\sum_{}^{}}}\mspace{2mu}\left( M_{i} - \overline{M} \right)^{2}}\sqrt{\overset{n}{\underset{i = 1}{\sum_{}^{}}}\mspace{2mu}\left( O_{i} - \overline{O} \right)^{2}}}
\label{eq:5}
\end{equation}

where \(\overline{M}\) and \(\overline{O}\) are the mean modeled and observed water levels, respectively. AutoCF additionally reports observed and modeled peak levels, peak-magnitude and peak-timing errors, and amplitude diagnostics. These metrics characterize overall error, systematic bias, temporal agreement, and event-peak behavior and are commonly used in CCF model evaluation \citep{lee2025,sebastian2021,vanormondt2025}.

\subsubsection{HWM evaluation}
\label{sec:2.5.3}

HWMs provide spatially distributed observations of event-maximum water level, complementing continuous gauge records in areas where permanent stations are sparse. AutoCF supports USGS Short-Term Network HWMs and user-supplied observations and can filter records using available quality classifications.

For each accepted HWM, AutoCF samples the simulated maximum water-surface elevation at the observation location. Under the default policy, quantitative residuals are calculated only when vertical-reference compatibility has been established through matching datums or an explicitly configured vertical offset. AutoCF records the datum relationship, any applied offset, and the distance over which the observation was matched to the model grid, reducing the risk that vertical-reference or spatial-matching differences are interpreted as hydraulic-model error.

\subsubsection{Evaluation products}
\label{sec:2.5.4}

Evaluation outputs include maximum-depth rasters, station hydrographs and metric tables, HWM comparisons, and summary reports. These products are retained for subsequent scientific analysis and presented through the AutoCF results workflow.

\subsection{Impact analysis and compound-flood attribution}
\label{sec:2.6}

AutoCF extends hydraulic simulation beyond flood depth mapping by quantifying exposure and separating the contributions of individual flood drivers to both hazard and consequences. Two complementary analysis modes are provided. Single-run analysis quantifies impacts for an individual simulation, while the attribution workflow uses a complete factorial experiment to separate the contributions and interactions of coastal, precipitation, and river forcing to hazard, exposure, and flood persistence. Here, flood persistence refers to the duration of flooding and is quantified using time-integrated metrics such as flooded area-hours and building-hours.

\subsubsection{Single-run exposure analysis}
\label{sec:2.6.1}

For a completed simulation, AutoCF derives flood footprints at user-defined depth thresholds and intersects them with building footprints and road features from Overture Maps and gridded population from WorldPop \citep{tatem2017}. At each threshold, the analysis calculates flooded area, affected buildings, the number of impacted road segments, flooded road length, and affected population. These quantities characterize the consequences of the selected simulation independently of driver attribution.

\subsubsection{Factorial driver scenarios}
\label{sec:2.6.2}

CCF attribution uses three driver groups: coastal forcing (\(C\)), precipitation (\(P\)), and river discharge (\(R\)). Coastal forcing combines the prescribed ocean water-level boundary with wind and atmospheric pressure, so \(C\) represents the integrated coastal forcing pathway.

For three drivers, AutoCF constructs the complete 2\textsuperscript{3} factorial set,

\[\{ NONE,C,P,R,CP,CR,PR,CPR\}\text{.}\]

The original full simulation is reused as \(CPR\), requiring seven additional simulations. Terrain, grid structure, surface properties, and numerical settings remain fixed across scenarios, while only the active forcing components change. Inactive gridded and point forcing are replaced by neutral conditions while preserving valid spatial and temporal structures. Scenario outputs are checked for consistent dimensions, coordinates, and coordinate reference systems before attribution. The complete factorial design retains information on both individual driver effects and nonlinear interactions among them.

\subsubsection{Exact Shapley attribution}
\label{sec:2.6.3}

Driver contributions are quantified using the Shapley value (Shapley, 1953). Let \(N = \left\{ C,P,R \right\}\) denote the full driver set and \(f(S)\) a hazard or consequence metric obtained when the subset \(S \subseteq N\) is active. The contribution of driver \(d\) is

\begin{equation}
\phi_{d} = \underset{S \subseteq N \smallsetminus \{ d\}}{\sum_{}^{}}\mspace{2mu}\frac{|S|!(|N| - |S| - 1)!}{|N|!}\left\lbrack f\left( S \cup \{ d\} \right) - f(S) \right\rbrack
\label{eq:6}
\end{equation}

Here, \(|S|\) denotes the number of drivers in subset \(S\), and \(|N|\) denotes the total number of drivers.

Because all eight driver combinations are simulated, the Shapley values are evaluated exactly. The method is applied spatially to maximum flood depth and to aggregated quantities including flooded area, affected buildings, affected population, flooded area-hours, and building-hours. Exact Shapley attribution satisfies

\begin{equation}
\underset{i \in N}{\sum_{}^{}}\mspace{2mu}\phi_{i} = f(N) - f(\varnothing)
\label{eq:7}
\end{equation}

so, the driver contributions reconstruct the difference between the full \(CPR\) simulation and the background \(NONE\) scenario. Negative contributions are retained because they represent suppressive or nonlinear effects and are required for exact closure. AutoCF also quantifies road exposure in its single-run impact analysis by reporting affected road segments and flooded road length. These road metrics are not included in the factorial analysis presented here; consequently, the exposure-attribution results in this study focus on buildings and population.

The primary attribution uses absolute flood depth for all factorial scenarios on a common land mask and therefore describes the transition from \(NONE\) to \(CPR\). AutoCF additionally calculates event-induced excess depth,

\begin{equation}
h_{S}^{exc} = \max\left( h_{S} - h_{NONE},0 \right)
\label{eq:8}
\end{equation}

where \(h_{S}\) is the maximum depth for scenario \(S\). The analysis mask remains fixed across scenarios, and dry model cells are represented as zero depth so that changes in attributed hazard or exposure are evaluated over a common domain.

\subsubsection{Factorial interactions and compound amplification}
\label{sec:2.6.4}

The full factorial experiment also permits explicit decomposition of non-additive interactions. Pairwise interactions are

\begin{equation}
I_{CP} = f(CP) - f(C) - f(P) + f(NONE)
\label{eq:9}
\end{equation}

\begin{equation}
I_{CR} = f(CR) - f(C) - f(R) + f(NONE)
\label{eq:10}
\end{equation}

\begin{equation}
I_{PR} = f(PR) - f(P) - f(R) + f(NONE)
\label{eq:11}
\end{equation}

and, the three-way interaction is

\begin{equation}
I_{CPR}=f(CPR)-f(CP)-f(CR)-f(PR)+f(C)+f(P)+f(R)-f(NONE)
\label{eq:12}
\end{equation}

Positive interaction values indicate enhancement relative to additive lower-order effects, while negative values indicate suppression.

AutoCF also calculates compound amplification from maximum flood depth as

\begin{equation}
A = h_{CPR} - \max\left( h_{C},h_{P},h_{R} \right)
\label{eq:13}
\end{equation}

Thus, \(A > 0\) identifies locations where combined forcing produces deeper flooding than any individual driver acting alone, whereas \(A < 0\) indicates that the combined simulation is shallower than the deepest single-driver simulation. Shapley values allocate the total event change among drivers, factorial interactions quantify departures from additivity, and compound amplification measures the signed depth difference between the combined event and the strongest single-driver realization.

\subsubsection{Exposure and duration attribution}
\label{sec:2.6.5}

The same factorial simulations propagate attribution from flood hazard to consequences. At each depth threshold, AutoCF calculates flooded land area, affected buildings, affected population, the corresponding \(NONE\) background, and the event-induced increment. Exact Shapley values are then calculated independently for each exposure metric. Population fractions use the population contained within a fixed analysis-land mask, which is held constant across all scenarios.

Time-varying outputs are additionally used to characterize flood persistence through flooded area-hours, building-hours, inundation duration, and the area or number of buildings exceeding user-defined persistence periods. Threshold-crossing times are linearly interpolated between saved model outputs. Duration attribution uses the existing factorial simulations and therefore requires no additional hydrodynamic scenarios.

\subsubsection{Social vulnerability and Driver Impact Shift}
\label{sec:2.6.6}

For U.S. applications, AutoCF combines the full-event flood footprint with the 2022 Centers for Disease Control and Prevention/Agency for Toxic Substances and Disease Registry Social Vulnerability Index (SVI). Analysis is performed at the census-tract scale and reports affected population by national SVI quartile, the affected fraction within each quartile, and a burden concentration ratio. The remaining hazard, exposure, and attribution analyses are independent of SVI availability and therefore remain applicable outside the U.S.

To determine whether the spatial footprint of a driver is proportional to its consequences, AutoCF introduces and calculates Driver Impact Shift (DIS). Let \(\phi_{d}^{(M)}\) denote the Shapley contribution of driver \(d \in \left\{ C,P,R \right\}\) to consequence metric \(M\), and let \(\phi_{d}^{(F)}\) denote its contribution to flooded area \(F\). The corresponding fractional Shapley shares are

\begin{equation}
s_{d}^{(M)} = \frac{\phi_{d}^{(M)}}{\underset{j \in \{ C,P,R\}}{\sum_{}^{}}\mspace{2mu}\phi_{j}^{(M)}}
\label{eq:14}
\end{equation}

and

\begin{equation}
s_{d}^{(F)} = \frac{\phi_{d}^{(F)}}{\underset{j \in \{ C,P,R\}}{\sum_{}^{}}\mspace{2mu}\phi_{j}^{(F)}}
\label{eq:15}
\end{equation}

DIS is then defined as the difference between these two shares,

\begin{equation}
{DIS}_{d,M} = 100\left( s_{d}^{(M)} - s_{d}^{(F)} \right)
\label{eq:16}
\end{equation}

or equivalently,

\begin{equation}
{DIS}_{d,M} = 100\left\lbrack \frac{\phi_{d}^{(M)}}{\underset{j}{\sum_{}^{}}\mspace{2mu}\phi_{j}^{(M)}} - \frac{\phi_{d}^{(F)}}{\underset{j}{\sum_{}^{}}\mspace{2mu}\phi_{j}^{(F)}} \right\rbrack
\label{eq:17}
\end{equation}

A positive \({DIS}_{d,M}\) indicates that driver \(d\) contributes a larger share of consequence \(M\) than of the flooded-area footprint. In contrast, a negative value indicates that the driver contributes a larger share of flooded area than of the corresponding consequence. Thus, DIS identifies shifts between the spatial distribution of flood hazard and the distribution of its consequences. For example, a driver may account for a relatively modest fraction of flooded area while contributing disproportionately to affected population or buildings because its inundation is concentrated in more highly exposed locations.

When the Shapley sums for both \(M\) and \(F\) are nonzero, the shares for each metric sum to unity and therefore

\begin{equation}
\underset{d \in \{ C,P,R\}}{\sum_{}^{}}\mspace{2mu}{DIS}_{d,M} = 0
\label{eq:18}
\end{equation}

so, DIS represents a redistribution of relative importance among the drivers rather than an additional physical contribution or interaction term.

\subsection{Reproducibility, deployment, and assisted interpretation}
\label{sec:2.7}

AutoCF treats reproducibility as a property of the complete CCF workflow. The project configuration, prepared datasets, provenance records, run state, scientific outputs, and software environment are retained so that model construction and subsequent analyses can be traced and repeated. This extends reproducibility beyond the final hydraulic input files to the data preparation, execution, evaluation, and attribution stages that produced the reported results.

\subsubsection{Project state and output organization}
\label{sec:2.7.1}

Stage-level caching, resumability, and dependency-aware invalidation follow the procedures described in Section~\ref{sec:2.4.4}. In addition, AutoCF organizes each project so that scientific inputs, model files, observations, factorial scenarios, exposure products, figures, tables, reports, logs, and metadata remain traceable to the corresponding run. Run metadata include the resolved configuration, execution state, and an artifact catalog linking generated products to their scientific role.

The Results interface uses this catalog to present the principal scientific products while retaining the underlying NetCDF, GeoTIFF, GIS, model, and log files required for reproducibility and further analysis. This separation keeps technical artifacts available without requiring users to navigate internal implementation files when interpreting model results.

\subsubsection{Cross-platform scientific execution}
\label{sec:2.7.2}

AutoCF uses a common project configuration and controlled scientific workflow across Windows, macOS, containerized environments, and high-performance computing (HPC) systems. The desktop interface is separated from the version-controlled geospatial, model-construction, and hydraulic software environment, reducing platform-dependent differences in scientific execution.

On Windows, the scientific environment can run through Windows Subsystem for Linux 2 (WSL2) or Docker, with validated CPU and compatible NVIDIA GPU SFINCS executables. macOS distributions provide native desktop applications for Intel and Apple-silicon systems while running the scientific stack in a containerized environment. Platform-specific runtime information is retained so that numerical agreement and computational performance can be evaluated explicitly rather than inferred from successful execution alone.

For HPC deployment, AutoCF separates network-dependent data preparation from compute-intensive hydraulic execution. Prepared inputs, the resolved configuration, provider metadata, runtime information, and the factorial-scenario manifest are fixed before work is submitted to a site scheduler. The HPC package is scheduler-neutral: it neither allocates resources nor generates scheduler directives. Factorial scenarios are independent and restart-safe at scenario boundaries, allowing users to distribute them across scheduler jobs when desired, while containerized execution uses Apptainer or Singularity \citep{kurtzer2017}. This structure supports reproducible transfer of the same scientific project from desktop systems to HPC resources.

\subsubsection{LLM-assisted guidance and interpretation}
\label{sec:2.7.3}

AutoCF includes Riva, a local LLM-assisted interface for workflow guidance, documentation retrieval, and interpretation of completed analyses. Riva combines deterministic retrieval and result extraction with an optional lightweight local language model. Its response workflow prioritizes curated AutoCF guidance, saved scientific outputs, and packaged SFINCS and HydroMT-SFINCS documentation, and can use Qwen3.5-0.8B through the llama.cpp inference engine to synthesize the retrieved information into natural-language responses. All inference is performed locally, avoiding dependence on an external LLM service.

Scientific quantities used by Riva, including evaluation metrics, Shapley contributions, factorial interactions, compound amplification, exposure, duration, DIS, and SVI outputs, are calculated deterministically by AutoCF before interpretation. Riva therefore operates as an interpretation layer over established numerical results and documentation, preserving reproducibility of the underlying scientific analysis. Its access is read-oriented with respect to project configuration and hydraulic execution, while resource management limits competition between language-model inference and computationally intensive simulations.

\section{Case study and experimental design}
\label{sec:3}

The experiments were designed to evaluate AutoCF as an integrated CCF modeling and analysis system rather than to revalidate the underlying hydrodynamic solver. SFINCS is an established reduced-complexity hydrodynamic model that has been independently evaluated and applied to CCF modeling in a range of environments \citep{leijnse2021,liang2026,sebastian2021}. The purpose of the present experiments is therefore different. We assess whether the datasets and model components selected, harmonized, and assembled automatically by AutoCF can produce a physically credible representation of a major CCF event; whether the resulting hazard fields can be evaluated consistently against independent observations; whether the attribution framework can translate the simulated hazard into driver-specific exposure and impact information; and whether the same workflow can be reproduced across different computing environments.

\subsection{Demonstration scope and benchmark selection}
\label{sec:3.1}

AutoCF is not restricted to a particular geographic region. Its globally applicable pathways include datasets for event identification, terrain, atmospheric forcing, coastal water levels, river discharge, land cover, and population. Examples include the IBTrACS, GTSM, GloFAS, GEDTM30, Copernicus DEM GLO-30, ERA5, WorldCover, GCN250, and WorldPop. These sources allow a CCF project to be constructed in many regions without requiring a predefined local database. AutoCF also accepts user-supplied terrain, atmospheric forcing, coastal water levels, river discharge, land-cover information, and other inputs so that higher-resolution or locally validated datasets can replace the built-in sources where available.

Global coverage, however, does not imply globally uniform data quality. High-resolution topography and bathymetry, long observational records, datum-controlled water-level measurements, river discharge observations, and event-specific validation data remain unevenly distributed. In many regions, a detailed application therefore requires locally acquired datasets to supplement or replace globally consistent products. The objective of this study is consequently not to demonstrate uniform predictive skill of the built-in global datasets everywhere. Instead, we use a data-rich benchmark to test the complete AutoCF workflow under conditions where its automatically assembled hydraulic representation can be evaluated independently and where its post-processing and attribution capabilities can be examined in detail.

Hurricane Harvey (2017) in the Houston-Galveston region of southeast Texas is selected for this purpose. Harvey produced severe precipitation, river flooding, and elevated coastal water levels, making it a well-documented example of pluvial-fluvial-coastal compound flooding \citep{lee2026,sebastian2021,vanormondt2025}. It was also the second-costliest tropical cyclone to impact the United States, based on Consumer Price Index (CPI)-adjusted damage estimates \citep{ncei2025}. The event is particularly suitable for evaluating AutoCF because extensive water-level observations, river discharge measurements, and surveyed HWMs are available across the affected region.

The U.S. setting also allows the experiment to exercise several data pathways that are difficult to test simultaneously in many other regions, including high-resolution coastal topobathymetry, dense hydrological and oceanographic observations, detailed soil information, building footprints, population data, and social-vulnerability information.

\subsection{Study domain and automated model construction}
\label{sec:3.2}

\begin{figure}[!htbp]
\centering
\includegraphics[width=\linewidth,height=0.7\textheight,keepaspectratio]{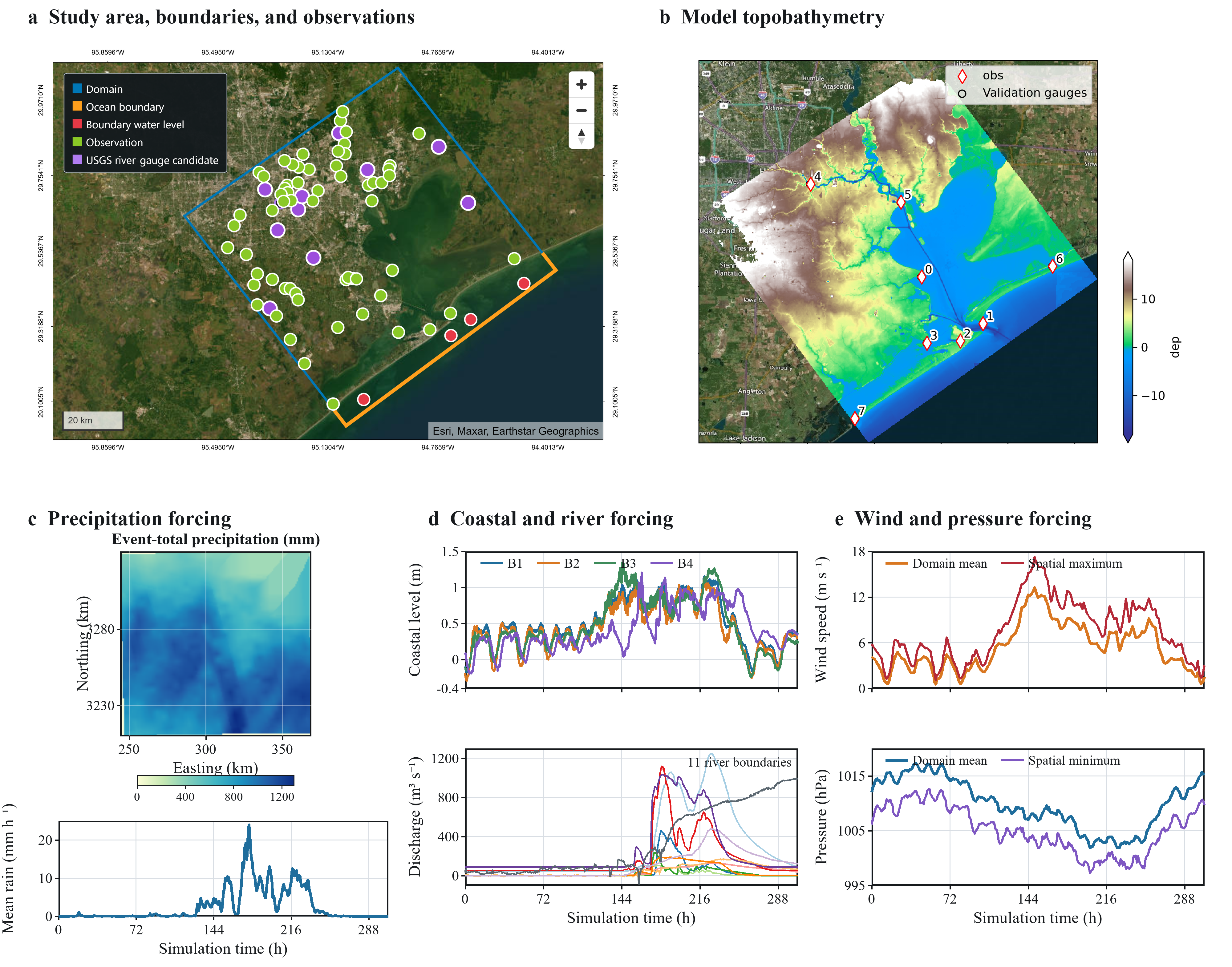}
\caption{Study area, model topobathymetry, and Hurricane Harvey forcing during the CCF simulation. (\textbf{a}) Geographic extent of the SFINCS domain, ocean boundary, prescribed water-level boundary locations, observations, and candidate USGS river gauges. (\textbf{b}) Model topobathymetry and locations used for water-level output and validation. (\textbf{c}) Spatial distribution of event-total precipitation and the corresponding domain-mean precipitation rate. The domain mean is the arithmetic average across all atmospheric-forcing grid cells at each simulation time. (\textbf{d}) Prescribed water levels at the four coastal boundary points and discharge hydrographs at the 11 river boundaries. B1--B4 follow the row order in SFINCS boundary setup; their west-to-east order is B3, B1, B2, and B4. (\textbf{e}) Evolution of wind speed and atmospheric pressure. For each time step, the domain-mean curves represent averages across the entire atmospheric grid, whereas the spatial maximum wind speed and spatial minimum pressure represent the most extreme grid-cell values anywhere in the domain.}
\label{fig:2}
\end{figure}

The study domain covers Galveston Bay, the Houston metropolitan area, the lower coastal plain, and the adjacent Gulf of Mexico shoreline (Figure~\ref{fig:2}). The domain extends approximately from 29.03 to 30.06° N and from 95.61 to 94.38° W. AutoCF constructed a 200 m UTM Zone 15N computational grid containing 595 × 571 cells. The resulting model contains 170,956 active water-level cells and 340,537 active velocity points. A SFINCS subgrid representation with 10 elevation levels derived from 20 × 20 fine-scale samples per computational cell was used to retain information on fine-scale terrain and conveyance while maintaining a computationally tractable regional model.

The simulation extends from 20 August 2017 00:00 UTC to 1 September 2017 18:00 UTC, providing a 306 h window that includes pre-event conditions, the principal rainfall and coastal water level response, and the subsequent recession. AutoCF assembled the terrain, surface properties, atmospheric forcing, coastal boundary conditions, river inflows, and observational datasets through the workflow described in Section~\ref{sec:2}. NOAA CUDEM topobathymetry was used as the primary terrain source, with GEBCO used for gap filling. The terrain was referenced to the North American Vertical Datum of 1988 (NAVD88). Surface roughness was derived from ESA WorldCover, while infiltration was represented using GCN250 Curve Number data for average antecedent moisture conditions.

Atmospheric forcing was provided by the NOAA AORC, including spatially distributed precipitation, wind, and atmospheric pressure. Coastal water-level forcing was obtained from NOAA CO-OPS, and river discharge was obtained from USGS observations selected through the AutoCF spatial workflow. Coastal forcing in the subsequent attribution experiment comprises the prescribed coastal water level boundary together with wind and atmospheric pressure, whereas precipitation and river discharge are treated as separate driver groups.

The model was generated through the same configuration-driven procedure available to AutoCF users. No automatic event-specific calibration was applied against the validation observations. The experiment therefore evaluates the hydraulic representation obtained from the assembled datasets and specified model configuration. Table~\ref{tab:1} summarizes the complete numerical configuration and data sources, while Figure~\ref{fig:2} shows the study domain, topobathymetry, observational locations, and principal forcing fields and time series.

\begin{table}[!htbp]
\centering
\caption{Hurricane Harvey benchmark model configuration, forcing, and attribution settings.}
\label{tab:1}
\begingroup
\small
\setlength{\tabcolsep}{4.0pt}
\renewcommand{\arraystretch}{1.18}
\begin{tabular}{@{}>{\raggedright\arraybackslash}p{\dimexpr 0.21951\linewidth-3.512pt\relax}>{\raggedright\arraybackslash}p{\dimexpr 0.44715\linewidth-7.154pt\relax}>{\raggedright\arraybackslash}p{\dimexpr 0.33333\linewidth-5.333pt\relax}@{}}
\toprule
\textbf{Component} & \textbf{Configuration} & \textbf{Reporting detail} \\
\midrule
Simulation period & 20 Aug 2017 00:00 to 1 Sep 2017 18:00 UTC & 306 h \\
Domain & Houston-Galveston coastal domain & 29.03-30.06 \textsuperscript{\ensuremath{^{\circ}}}N\par 95.61-94.38 \textsuperscript{\ensuremath{^{\circ}}}W \\
Grid & Regular UTM Zone 15N grid & 200 m resolution\par (595 x 571 cells) \\
Active mesh & 170,956 water-level cells & 340,537 velocity points \\
Subgrid representation & 10 elevation levels & 20 x 20 fine-scale\par samples per cell \\
Topobathymetry & NOAA CUDEM 1/9 arc-sec with GEBCO gap filling & Target datum NAVD88 \\
Roughness & ESA WorldCover 2021 at 10 m & Class-based Manning coefficients \\
Infiltration & Curve Number method & GCN250 and SSURGO; average antecedent moisture \\
Atmospheric forcing & AORC precipitation, wind, and pressure & Spatially distributed NetCDF fields \\
Coastal boundary & NOAA CO-OPS water levels & NAVD88-referenced open boundary \\
River forcing & USGS discharge observations & Automatically selected inflow stations \\
Hydrodynamics & Subgrid SFINCS with advection, viscosity, wind, pressure, rainfall, and infiltration & SFINCS v2.4.0 Galibier \\
Numerical controls & alpha=0.5; theta=1.0;\par huthresh=0.05 m & Maximum time step 60 s\par Mean time step 7.328 s \\
Attribution & Eight coastal-precipitation-river factorial scenarios & Depth thresholds 0.15, 0.30, 0.60, 1.0, and 2.0 m \\
\bottomrule
\end{tabular}
\endgroup
\end{table}

\subsection{Observational evaluation design}
\label{sec:3.3}

The first experimental objective is to determine whether the automated preprocessing and model-construction workflow produces a credible representation of the observed CCF event. This evaluation tests the complete AutoCF modeling chain, including terrain preparation, surface characterization, boundary selection, forcing harmonization, model construction, execution, vertical-datum handling, and observational post-processing.

Continuous water-level observations and event HWMs provide complementary tests. Water-level hydrographs evaluate the temporal evolution, magnitude, phase, and peak response across coastal and inland locations. Evaluation follows the metrics defined in Section~\ref{sec:2.5}, including RMSE, MAE, bias, correlation, peak-magnitude error, and peak timing. HWMs provide a spatially distributed test of event-maximum water-surface elevation at overland locations where continuous NOAA gauge observations are unavailable.

\subsection{Compound-flood impact and attribution experiment}
\label{sec:3.4}

The second experimental objective is to determine whether AutoCF can move beyond flood simulation and convert hydrodynamic output into interpretable information about the sources and consequences of a CCF event. The Harvey experiment therefore applies the complete coastal-precipitation-river factorial framework described in Section~\ref{sec:2.6}.

Three driver groups are considered: coastal forcing \(C\), precipitation \(P\), and river discharge \(R\). The full \(2^{3}\) factorial experiment consists of \emph{NONE}, \emph{C}, \emph{P}, \emph{R}, \emph{CP}, \emph{CR}, \emph{PR}, and \emph{CPR}. The original compound simulation is reused as CPR, requiring seven additional hydrodynamic simulations. All scenarios retain the same terrain, grid, surface characteristics, analysis domain, and numerical configuration, with only the active forcing components changed. This design allows both individual driver contributions and nonlinear interactions among the drivers to be resolved.

Attribution is evaluated at flood-depth thresholds of 0.15, 0.30, 0.60, 1.0, and 2.0 m. Exact Shapley values are calculated for flood hazard and for consequence metrics including flooded area, exposed buildings, exposed population, flooded area-hours, and building-hours. The complete factorial experiment is also used to quantify pairwise and three-way interactions and compound amplification. For the U.S. benchmark, the affected population is additionally summarized by national quartile of the 2022 CDC/ATSDR Social Vulnerability Index (SVI) at the census-tract level.

Particular attention is given to the newly introduced DIS metric, which compares the Shapley share of a driver in a consequence metric with its Shapley share of flooded area. It therefore tests whether the spatial footprint associated with a particular flood driver translates proportionally into impacts. A positive DIS indicates that a driver is relatively more important for consequences than for flooded-area extent, whereas a negative value identifies a driver whose hazard footprint is proportionally larger than its societal consequence. The Harvey experiment consequently tests whether AutoCF can carry driver attribution from hydrodynamic hazard through exposure, persistence, and social consequence.

\subsection{Cross-platform reproducibility and computational experiment}
\label{sec:3.5}

The final experimental objective is to evaluate whether the AutoCF scientific workflow can be transferred across computing environments without materially changing its numerical results. An equivalent Harvey model setup was therefore executed on Windows, macOS, and a high-performance computing (HPC) system. The Windows implementation used the controlled Linux scientific environment through Windows Subsystem for Linux 2, the macOS implementation used the containerized scientific environment on Apple silicon, and the HPC implementation used the validated GPU-enabled SFINCS runtime. The HPC simulation was executed on the Princeton Della cluster using an NVIDIA A100 80 GB PCIe GPU configured with a 3g.40gb Multi-Instance GPU (MIG) profile, providing approximately 40 GB of GPU memory.

The model domain, core hydraulic settings, forcing, and prepared scientific inputs were kept consistent across the platform experiments; platform-specific runtime and output-control settings were recorded separately. Reproducibility was assessed by comparing the core configuration and prepared inputs, water-level hydrographs, and event-maximum water levels calculated from the common hourly map outputs. Computational performance was evaluated separately using elapsed runtime and solver timing information. The HPC/GPU solution was used as the reference realization for the full hazard, evaluation, and factorial-attribution analysis because the eight-scenario attribution experiment requires repeated hydrodynamic simulation and the HPC backend provides the computational throughput required for this experiment. Windows and macOS executions are used to evaluate portability and numerical reproducibility.

Together, these experiments test three distinct aspects of AutoCF: the reliability of automated CCF model construction and evaluation, the ability to transform hydrodynamic output into driver-specific impact information, and the portability and numerical reproducibility of hydrodynamic execution across computing environments.

\section{Results}
\label{sec:4}

The Hurricane Harvey benchmark was used to evaluate three aspects of AutoCF: the compound-flood simulation assembled by the automated workflow, its agreement with independent observations, and the ability of the framework to translate simulated flooding into driver-specific hazard and impact information. Cross-platform experiments are additionally used to assess whether the same modelling workflow produced consistent results across computing environments. The full CCF simulation provides the reference state for the observational evaluation and for the subsequent factorial attribution analysis.

\subsection{CCF simulation and exposure}
\label{sec:4.1}

\begin{figure}[!htbp]
\centering
\includegraphics[width=\linewidth,height=0.76\textheight,keepaspectratio]{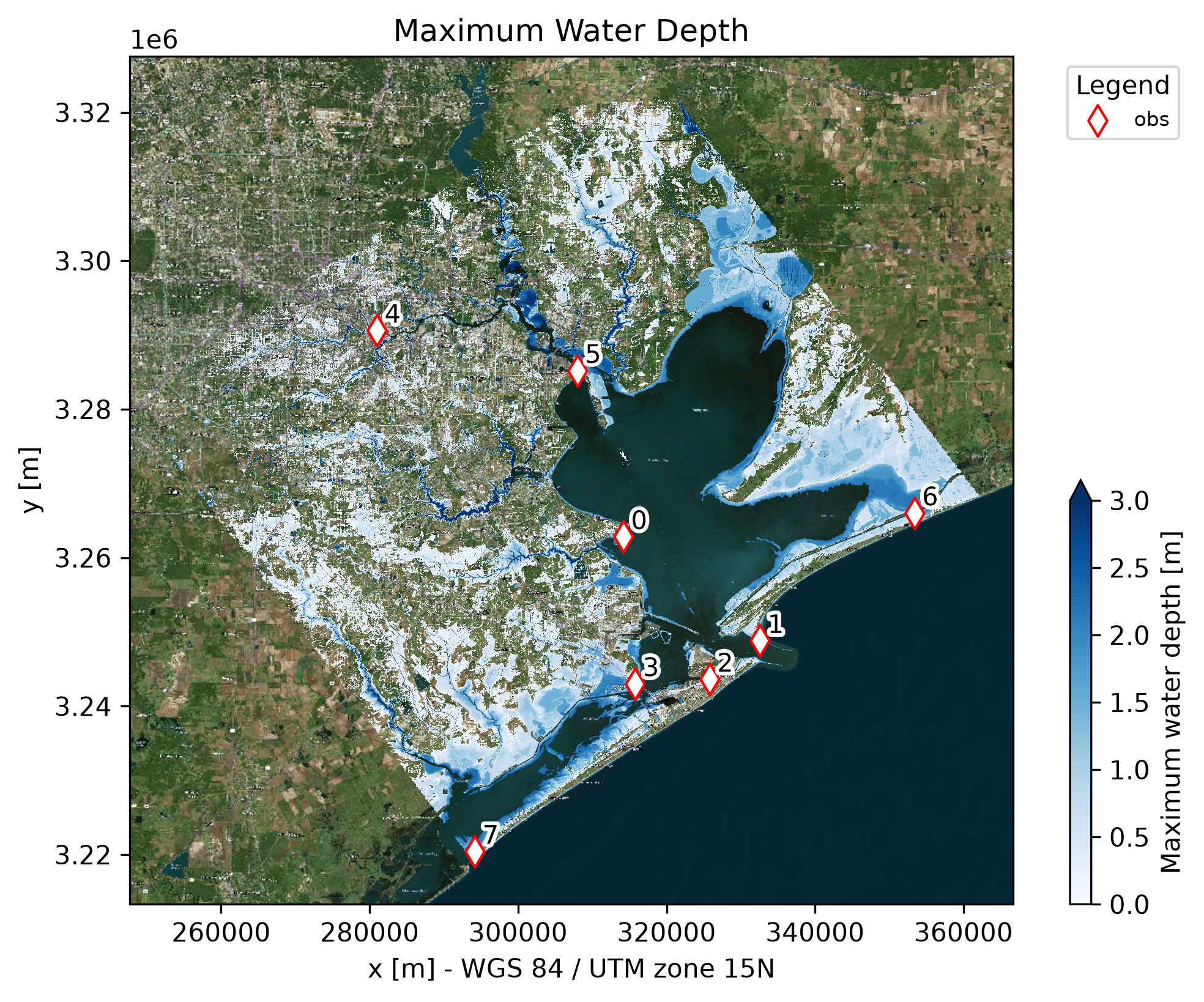}
\caption{Maximum simulated water depth for Hurricane Harvey's full CPR scenario. Validation locations are shown for reference. These gauge locations are numbered as follows: 0, Eagle Point, Galveston Bay; 1, Galveston Bay Entrance, North Jetty; 2, Galveston Pier 21; 3, Galveston Railroad Bridge; 4, Manchester; 5, Morgans Point, Barbours Cut; 6, Rollover Pass; and 7, San Luis Pass. Gauge names correspond to those listed in Table~\ref{tab:3}.}
\label{fig:3}
\end{figure}

The full CPR simulation produced widespread inundation across the Houston-Galveston domain (Figure~\ref{fig:3}). Flooding extended from the Gulf shoreline and Galveston Bay into river corridors, drainage networks, and inland low-lying areas. Deeper inundation was concentrated along channels and topographic depressions, while shallower flooding covered a substantially larger portion of the coastal plain.

The simulated flooded area decreased by 53.8\%, from 6,033.68 km² at the 0.15 m threshold to 2,790.72 km² at 2.0 m (Table~\ref{tab:2}). Exposure showed a similar depth dependence. At 0.15 m, the flood footprint intersected 694,616 buildings and approximately 1.84 million people. At 1.0 m, building and population exposure had decreased by 72.3\% and 70.8\%, respectively, to 192,387 buildings and approximately 535,000 people. At 2.0 m, the corresponding reductions reached 85.9\% and 84.6\%, leaving 97,651 buildings and approximately 282,000 people exposed.

\begin{table}[!htbp]
\centering
\caption{Simulated flood extent and exposure for the full CPR scenario across five water-depth thresholds.}
\label{tab:2}
\begingroup
\small
\setlength{\tabcolsep}{4.0pt}
\renewcommand{\arraystretch}{1.18}
\begin{tabular}{@{}>{\raggedright\arraybackslash}p{\dimexpr 0.16260\linewidth-5.203pt\relax}>{\raggedright\arraybackslash}p{\dimexpr 0.19512\linewidth-6.244pt\relax}>{\raggedright\arraybackslash}p{\dimexpr 0.20325\linewidth-6.504pt\relax}>{\raggedright\arraybackslash}p{\dimexpr 0.21951\linewidth-7.024pt\relax}>{\raggedright\arraybackslash}p{\dimexpr 0.21951\linewidth-7.024pt\relax}@{}}
\toprule
\textbf{Depth threshold (m)} & \textbf{Flooded area (km\textsuperscript{2})} & \textbf{Buildings exposed} & \textbf{Population exposed} & \textbf{Population affected (\%)} \\
\midrule
0.15 & 6,033.68 & 694,616 & 1,835,688 & 84.9 \\
0.30 & 5,284.76 & 527,323 & 1,434,053 & 66.3 \\
0.60 & 4,380.16 & 303,791 & 838,162 & 38.8 \\
1.00 & 3,756.72 & 192,387 & 535,215 & 24.7 \\
2.00 & 2,790.72 & 97,651 & 282,241 & 13.1 \\
\bottomrule
\end{tabular}
\endgroup
\end{table}

Figure~\ref{fig:3} and Table~\ref{tab:2} describe the absolute hazard and exposure produced by the full event. These results are used in two different ways below. Section~\ref{sec:4.2} evaluates whether the simulated water levels are consistent with observations, while Section~\ref{sec:4.4} separates the same compound event into coastal, precipitation, and river contributions.

\subsection{Observational evaluation}
\label{sec:4.2}

AutoCF evaluation was performed independently using continuous NOAA water-level records and USGS HWMs. The two observation types provide complementary evidence. Gauge hydrographs evaluate the temporal evolution of water levels, whereas HWMs provide spatial information on event-maximum elevation.

\subsubsection{Water-level hydrographs}
\label{sec:4.2.1}

\begin{figure}[!htbp]
\centering
\includegraphics[width=\linewidth,height=0.76\textheight,keepaspectratio]{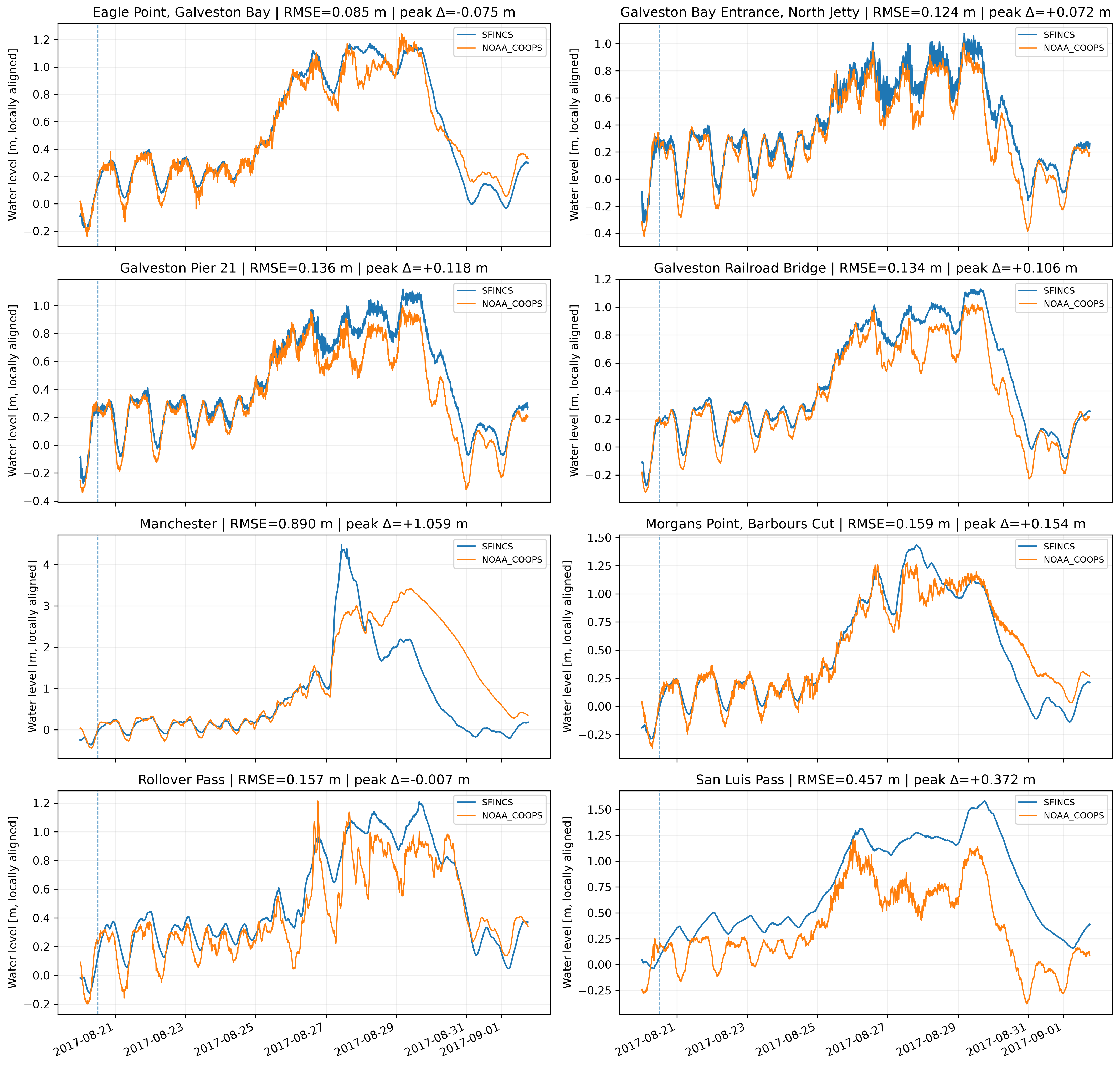}
\caption{Observed and simulated water-level hydrographs at eight NOAA CO-OPS stations. The dashed line identifies the beginning of the common evaluation period. Station RMSE and peak-magnitude error are reported in each panel.}
\label{fig:4}
\end{figure}

The simulated hydrographs reproduce the principal event-scale water-level variability at the eight NOAA CO-OPS stations shown in Figure~\ref{fig:4}. Across the stations, the median RMSE is 0.147 m, the median MAE is 0.108 m, and the median correlation is 0.951. Six stations have RMSE values between 0.085 and 0.159 m, with peak-magnitude errors ranging from -0.075 to +0.154 m (Table~\ref{tab:3}).

Agreement is strongest at most coastal and bay stations, while larger errors occur at Manchester and San Luis Pass. At Manchester, the model captures the rapid rise but overestimates the event peak and does not reproduce the prolonged observed recession, producing an RMSE of 0.890 m and a correlation of 0.741. At San Luis Pass, the RMSE is 0.457 m and the correlation is 0.823. This gauge (site 7 in Figure~\ref{fig:3}) lies immediately adjacent to the model's open boundary, leaving insufficient spatial buffer to resolve tidal and surge propagation and local exchange through the pass; the model therefore cannot fully reproduce the observed water-level dynamics at this station. The largest peak-time differences occur at San Luis Pass and Rollover Pass, where several tidal peaks have similar magnitudes.

The combined results in Figure~\ref{fig:4} and Table~\ref{tab:3} therefore show strong agreement over most of the coastal and estuarine domain, while also identifying locations where the regional configuration produces larger local discrepancies.

\begin{table}[!htbp]
\centering
\caption{Water-level evaluation of the Hurricane Harvey simulation against eight NOAA CO-OPS stations. Positive bias and peak error indicate model overprediction, and positive peak lag indicates a later simulated peak.}
\label{tab:3}
\begingroup
\footnotesize
\setlength{\tabcolsep}{2.0pt}
\renewcommand{\arraystretch}{1.18}
\begin{tabular}{@{}>{\raggedright\arraybackslash}p{\dimexpr 0.26500\linewidth-8.480pt\relax}>{\raggedright\arraybackslash}p{\dimexpr 0.09187\linewidth-2.940pt\relax}>{\raggedright\arraybackslash}p{\dimexpr 0.09187\linewidth-2.940pt\relax}>{\raggedright\arraybackslash}p{\dimexpr 0.09187\linewidth-2.940pt\relax}>{\raggedright\arraybackslash}p{\dimexpr 0.09187\linewidth-2.940pt\relax}>{\raggedright\arraybackslash}p{\dimexpr 0.09187\linewidth-2.940pt\relax}>{\raggedright\arraybackslash}p{\dimexpr 0.09187\linewidth-2.940pt\relax}>{\raggedright\arraybackslash}p{\dimexpr 0.09187\linewidth-2.940pt\relax}>{\raggedright\arraybackslash}p{\dimexpr 0.09187\linewidth-2.940pt\relax}@{}}
\toprule
\textbf{Station} & \textbf{RMSE (m)} & \textbf{MAE (m)} & \textbf{Bias (m)} & \textbf{r} & \textbf{Obs. peak (m)} & \textbf{Model peak (m)} & \textbf{Peak error (m)} & \textbf{Peak lag (h)} \\
\midrule
Eagle Point, Galveston Bay & 0.085 & 0.066 & 0.015 & 0.979 & 1.246 & 1.171 & -0.075 & -21.5 \\
Galveston Bay Entrance, North Jetty & 0.124 & 0.096 & 0.091 & 0.963 & 1.005 & 1.077 & 0.072 & 0.4 \\
Galveston Pier 21 & 0.136 & 0.104 & 0.100 & 0.961 & 0.999 & 1.117 & 0.118 & 0.3 \\
Galveston Railroad Bridge & 0.134 & 0.103 & 0.100 & 0.970 & 1.022 & 1.128 & 0.106 & 5.8 \\
Manchester & 0.890 & 0.566 & -0.373 & 0.741 & 3.414 & 4.473 & 1.059 & -47.4 \\
Morgans Point, Barbours Cut & 0.159 & 0.112 & -0.020 & 0.942 & 1.280 & 1.434 & 0.154 & 6.2 \\
Rollover Pass & 0.157 & 0.127 & 0.090 & 0.920 & 1.215 & 1.208 & -0.007 & 69.2 \\
San Luis Pass & 0.457 & 0.385 & 0.384 & 0.823 & 1.211 & 1.583 & 0.372 & 89.5 \\
\bottomrule
\end{tabular}
\endgroup
\end{table}

\textbf{Note:} Aggregate metrics based on all these eight NOAA gauges are: mean RMSE=0.268 m; mean MAE=0.195 m; mean bias=+0.048 m; median correlation=0.951; mean absolute peak error=0.245 m.

\subsubsection{USGS HWMs}
\label{sec:4.2.2}

The event-maximum simulated water surface was also evaluated against the USGS Harvey high-water-mark dataset. Of the 56 marks retained within the study area, 55 satisfied the 1 km sampling and NAVD88 compatibility criteria used by AutoCF. All metric-eligible records were classified as Excellent by USGS.

\begin{figure}[!htbp]
\centering
\includegraphics[width=\linewidth,height=0.76\textheight,keepaspectratio]{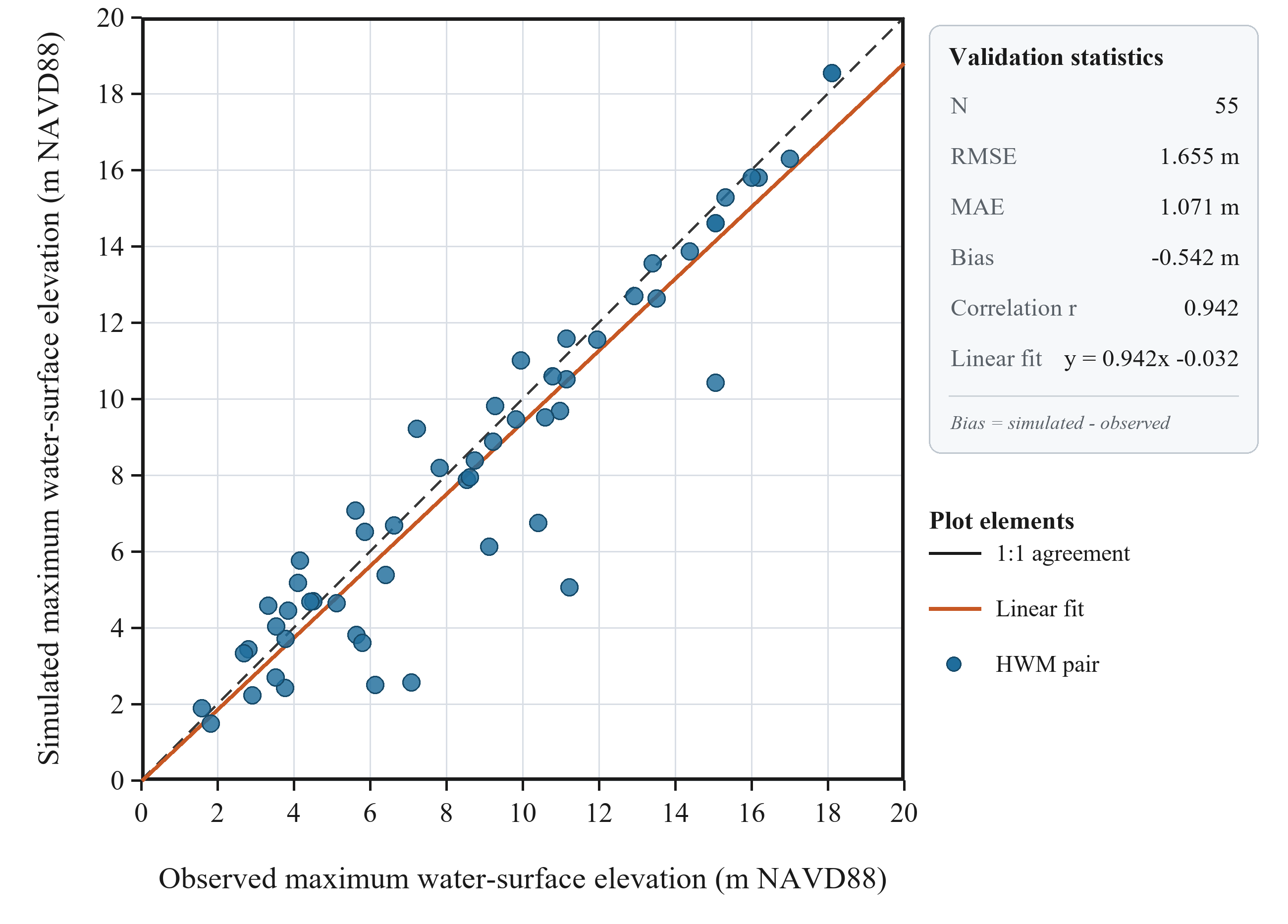}
\caption{Observed and simulated maximum water-surface elevations for 55 metric-eligible USGS HWMs. The figure shows the 1:1 line, fitted regression, and summary validation statistics.}
\label{fig:5}
\end{figure}

Observed and simulated elevations are strongly correlated over the 1.58-18.11 m observed range, with \emph{r} = 0.942 and a regression slope of 0.942 (Figure~\ref{fig:5}). The HWM RMSE is 1.655 m, the MAE is 1.071 m, and the mean bias is -0.542 m. The median absolute error is lower, at 0.628 m, and 67.3\% of the evaluated marks are reproduced within 1.0 m (Table~\ref{tab:4}).

The HWM results complement the hydrograph evaluation in Section~\ref{sec:4.2.1}. The gauge comparison primarily tests temporal evolution at fixed stations, whereas Figure~\ref{fig:5} shows that the automated model also retains the broad spatial variation in Harvey peak water levels across coastal and inland locations.

\begin{table}[!htbp]
\centering
\caption{Evaluation of simulated event-maximum water-surface elevation against USGS Hurricane Harvey HWMs.}
\label{tab:4}
\begingroup
\small
\setlength{\tabcolsep}{4.0pt}
\renewcommand{\arraystretch}{1.18}
\begin{tabular}{@{}>{\raggedright\arraybackslash}p{\dimexpr 0.29839\linewidth-4.774pt\relax}>{\raggedright\arraybackslash}p{\dimexpr 0.20161\linewidth-3.226pt\relax}>{\raggedright\arraybackslash}p{\dimexpr 0.50000\linewidth-8.000pt\relax}@{}}
\toprule
\textbf{Metric} & \textbf{Value} & \textbf{Definition} \\
\midrule
Selected HWMs & 56 & USGS Harvey marks in study-area record \\
Metric-eligible HWMs & 55 & Within 1 km; NAVD88 compatible \\
Correlation & 0.942 & Observed versus simulated elevation \\
Regression slope & 0.942 & Intercept=-0.032 m \\
RMSE & 1.655 m & Peak water-surface elevation \\
MAE & 1.071 m & Peak water-surface elevation \\
Mean bias & -0.542 m & Model minus observation \\
Median absolute error & 0.628 m & Robust central error \\
Within 0.50 m & 43.6\% & 24 of 55 marks \\
Within 1.00 m & 67.3\% & 37 of 55 marks \\
\bottomrule
\end{tabular}
\endgroup
\end{table}

\subsection{Cross-platform reproducibility and computational performance}
\label{sec:4.3}

An equivalent Harvey model setup was executed on Windows, macOS, and the HPC environment described in Section~\ref{sec:3.5} (Table~\ref{tab:5}). The prepared Windows and HPC model input files used by the solver were hash-identical based on 256-bit Secure Hash Algorithm (SHA-256) checks. Numerical agreement was then evaluated directly from the simulated outputs rather than inferred from identical configuration files.

\begin{table}[!htbp]
\centering
\caption{Computational environments and elapsed runtime for the equivalent Hurricane Harvey model setups.}
\label{tab:5}
\begingroup
\small
\setlength{\tabcolsep}{4.0pt}
\renewcommand{\arraystretch}{1.18}
\begin{tabular}{@{}>{\raggedright\arraybackslash}p{\dimexpr 0.12598\linewidth-5.039pt\relax}>{\raggedright\arraybackslash}p{\dimexpr 0.22835\linewidth-9.134pt\relax}>{\raggedright\arraybackslash}p{\dimexpr 0.32426\linewidth-12.970pt\relax}>{\raggedright\arraybackslash}p{\dimexpr 0.12456\linewidth-4.983pt\relax}>{\raggedright\arraybackslash}p{\dimexpr 0.11024\linewidth-4.409pt\relax}>{\raggedright\arraybackslash}p{\dimexpr 0.08661\linewidth-3.465pt\relax}@{}}
\toprule
\textbf{Platform} & \textbf{Host} & \textbf{Execution resources} & \textbf{Elapsed time (s)} & \textbf{Elapsed time (min)} & \textbf{Speedup} \\
\midrule
Windows CPU & Dell XPS 13 9350; Intel Core Ultra 7 258V & WSL2; 8 CPU threads; about 15.4 GiB runtime memory & 3,016.8 & 50.28 & 1.000 \\
macOS CPU & Apple M1 Mac mini & ARM64 Docker; 8 CPU threads; about 3.83 GiB container memory & 1,700.7 & 28.34 & 1.774 \\
HPC GPU & NVIDIA A100\par 80 GB PCIe & 3g.40gb MIG instance (\textasciitilde40 GB GPU memory); SFINCS GPU backend; 1 host thread & 571.1 & 9.52 & 5.282 \\
\bottomrule
\end{tabular}
\endgroup
\end{table}

At the domain scale, the CPU and GPU event-maximum water-level fields are very similar (Table~\ref{tab:6}). Across 170,530 jointly valid cells, the spatial RMSE is 0.042 m, the MAE is 0.003 m, and the mean CPU-minus-GPU difference is +0.002 m. A total of 98.80\% of cells differ by less than 0.01 m, 99.56\% by less than 0.05 m, and 99.64\% by less than 0.10 m. Only 48 cells differ by more than 1.0 m.

The gauge hydrographs show similarly close agreement between operating environments. At the 307 common hourly output times, the aggregate Windows-macOS RMSE is 0.010 m and the MAE is 0.0027 m. Seven stations have cross-platform RMSE values no greater than 0.0105 m. Manchester again shows the largest sensitivity, consistent with its larger observational error in Section~\ref{sec:4.2.1}.

\begin{table}[!htbp]
\centering
\caption{Cross-platform numerical agreement and observational evaluation for the equivalent Hurricane Harvey model setups.}
\label{tab:6}
\medskip\par\textbf{\textbf{A. Gauge-validation summary}}\par\smallskip
\begingroup
\small
\setlength{\tabcolsep}{4.0pt}
\renewcommand{\arraystretch}{1.18}
\begin{tabular}{@{}>{\raggedright\arraybackslash}p{\dimexpr 0.25000\linewidth-8.000pt\relax}>{\raggedright\arraybackslash}p{\dimexpr 0.16129\linewidth-5.161pt\relax}>{\raggedright\arraybackslash}p{\dimexpr 0.16129\linewidth-5.161pt\relax}>{\raggedright\arraybackslash}p{\dimexpr 0.16129\linewidth-5.161pt\relax}>{\raggedright\arraybackslash}p{\dimexpr 0.26613\linewidth-8.516pt\relax}@{}}
\toprule
\textbf{Metric} & \textbf{Windows CPU} & \textbf{macOS CPU} & \textbf{HPC GPU} & \textbf{Interpretation} \\
\midrule
Station-history interval & 6 min & 60 min & 6 min & macOS report contains hourly records \\
Mean RMSE (m) & 0.275 & 0.278 & 0.268 & HPC GPU lowest \\
Mean MAE (m) & 0.199 & 0.203 & 0.198 & HPC GPU lowest \\
Median correlation & 0.951 & 0.950 & 0.951 & Equivalent across platforms \\
Mean absolute peak error (m) & 0.348 & 0.344 & 0.245 & HPC GPU lowest \\
Mean absolute bias (m) & 0.142 & 0.148 & 0.147 & Windows CPU lowest \\
Manchester RMSE (m) & 0.945 & 0.939 & 0.889 & HPC GPU lowest \\
Manchester model peak (m) & 5.298 & 5.176 & 4.473 & Observed peak approximately 3.41 m \\
\bottomrule
\end{tabular}
\endgroup
\medskip\par\textbf{\textbf{B. Direct numerical agreement}}\par\smallskip
\begingroup
\small
\setlength{\tabcolsep}{4.0pt}
\renewcommand{\arraystretch}{1.18}
\begin{tabular}{@{}>{\raggedright\arraybackslash}p{\dimexpr 0.33210\linewidth-5.314pt\relax}>{\raggedright\arraybackslash}p{\dimexpr 0.23185\linewidth-3.710pt\relax}>{\raggedright\arraybackslash}p{\dimexpr 0.43604\linewidth-6.977pt\relax}@{}}
\toprule
\textbf{Comparison} & \textbf{Result} & \textbf{Interpretation} \\
\midrule
Prepared Windows-HPC inputs & All inputs were hash-identical & Transfer preserved the active model input set \\
Windows CPU versus macOS CPU gauges & Shared-hour RMSE=0.010 m & Seven station RMSE values \textless=0.0105 m; Manchester=0.0259 m \\
CPU versus HPC GPU spatial RMSE & 0.0422 m & Event-maximum water level \\
CPU versus HPC GPU spatial MAE & 0.00310 m & Event-maximum water level \\
Cells within 0.01 m & 98.801\% & 2,045 of 170,530 cells exceed 0.01 m \\
Cells within 0.05 m & 99.562\% & 746 cells exceed 0.05 m \\
Cells within 0.10 m & 99.643\% & 609 cells exceed 0.10 m \\
Cells differing by more than 1.0 m & 48 & 0.028\% of jointly valid cells \\
Seven gauge hydrographs & Inter-run RMSE \textless=0.003 m & Numerically indistinguishable at reporting scale \\
Manchester hydrograph & Inter-run RMSE=0.175 m & Localized backend sensitivity; GPU peak closer to observation \\
\bottomrule
\end{tabular}
\endgroup
\end{table}

This spatial concentration is visible in Figure 6a-b. Most of the domain shows negligible CPU-GPU differences, whereas the largest deviations are concentrated near Manchester. The same location was also the largest outlier in the observational comparison, linking the cross-platform experiment to the evaluation results rather than treating it as a separate software benchmark.

Execution time differed substantially among the three environments. The Harvey simulation required 50.28 min on Windows/WSL2, 28.34 min on macOS/Docker, and 9.52 min on the HPC/GPU backend (Table~\ref{tab:5}). The HPC execution was therefore 5.28 times faster than Windows and 2.98 times faster than macOS. However, 89.1\% of the HPC elapsed time was associated with output writing, compared with 39.7 s spent on the momentum and continuity calculations. For this configuration, model output rather than hydrodynamic computation therefore dominates the remaining HPC runtime.

\begin{figure}[!htbp]
\centering
\includegraphics[width=\linewidth,height=0.76\textheight,keepaspectratio]{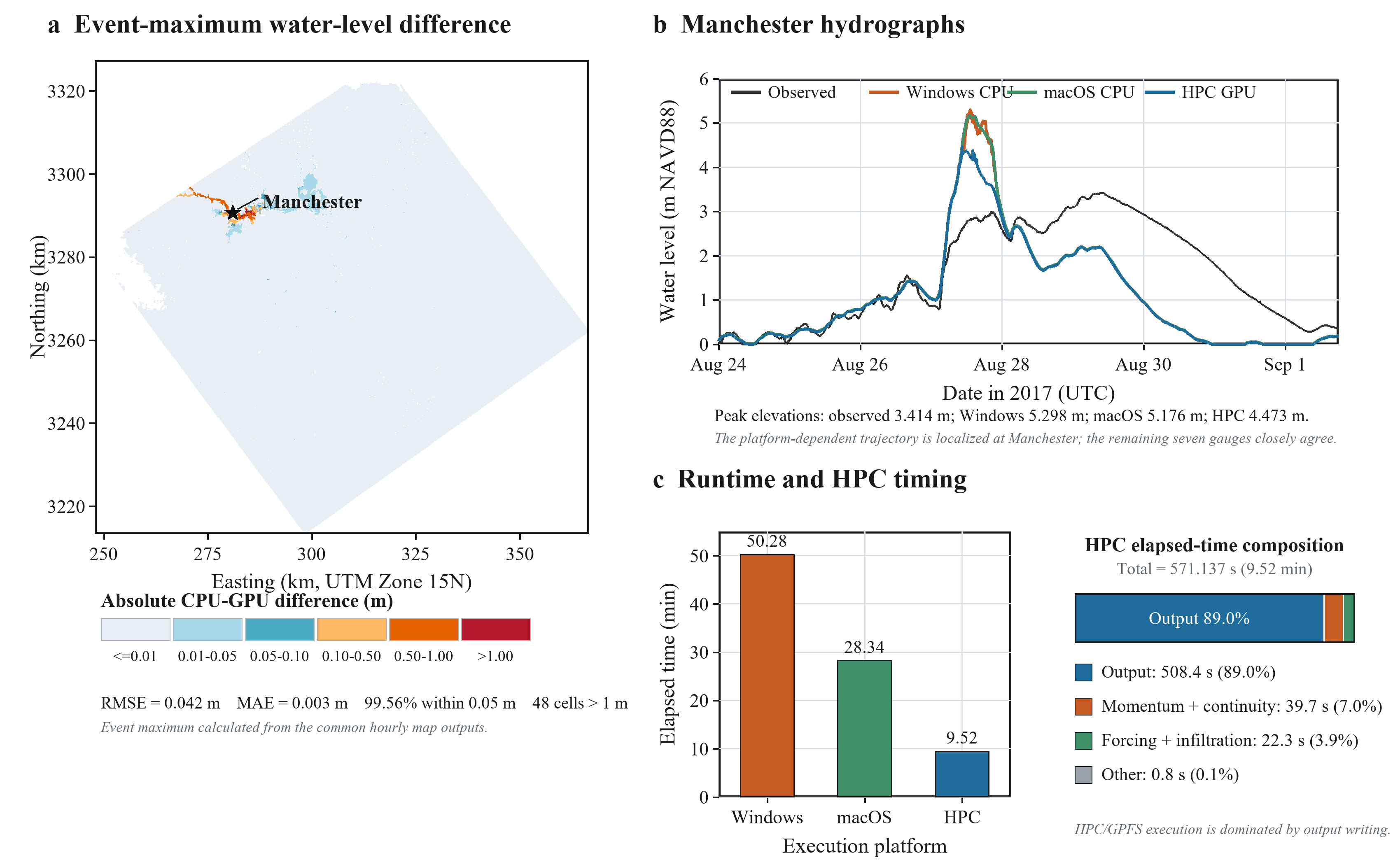}
\caption{Cross-platform reproducibility and performance. (\textbf{a}) Absolute difference in event-maximum water level between CPU and GPU solutions. (\textbf{b}) Hydrograph comparison at Manchester, where the largest localized backend sensitivity occurs. (\textbf{c}) Total runtime across computing environments and HPC elapsed-time composition.}
\label{fig:6}
\end{figure}

These results show that differences in hardware and runtime environment have little effect on the regional flood solution, while also revealing a localized area of numerical sensitivity. The HPC/GPU realization is therefore used for the factorial experiment below because it provides the required computational efficiency while remaining consistent with the broader cross-platform solution.

\subsection{CCF attribution and impacts}
\label{sec:4.4}

The validated full-event simulation was next decomposed into coastal \(C\), precipitation \(P\), and river \(R\) contributions using the complete factorial experiment described in Section~\ref{sec:3.4}. The analysis progresses from absolute scenario impacts to Shapley attribution, nonlinear interactions, duration, and the relationship between hazard footprint and societal consequences.

\subsubsection{Factorial scenarios and absolute impacts}
\label{sec:4.4.1}

The eight factorial scenarios show large differences in both hazard and exposure (Table~\ref{tab:7}). At the 0.30 m threshold, precipitation alone produces 5,156.44 km² of inundation, 513,846 exposed buildings, and approximately 1.405 million exposed people. The full \(C + P + R\) scenario increases these values to 5,284.76 km², 527,323 buildings, and approximately 1.434 million people.

\begin{table}[!htbp]
\centering
\caption{Flooded area, building exposure, and population exposure for the eight coastal-precipitation-river factorial scenarios at the 0.30 m depth threshold.}
\label{tab:7}
\begingroup
\small
\setlength{\tabcolsep}{4.0pt}
\renewcommand{\arraystretch}{1.18}
\begin{tabular}{@{}>{\raggedright\arraybackslash}p{\dimexpr 0.11161\linewidth-3.572pt\relax}>{\raggedright\arraybackslash}p{\dimexpr 0.21112\linewidth-6.756pt\relax}>{\raggedright\arraybackslash}p{\dimexpr 0.20500\linewidth-6.560pt\relax}>{\raggedright\arraybackslash}p{\dimexpr 0.21610\linewidth-6.915pt\relax}>{\raggedright\arraybackslash}p{\dimexpr 0.25617\linewidth-8.197pt\relax}@{}}
\toprule
\textbf{Scenario} & \textbf{Flooded area (km\textsuperscript{2})} & \textbf{Buildings exposed} & \textbf{Population exposed} & \textbf{Population affected (\%)} \\
\midrule
NONE & 2,483.80 & 20,647 & 31,158 & 1.4 \\
C & 3,153.28 & 39,109 & 64,321 & 3.0 \\
P & 5,156.44 & 513,846 & 1,404,940 & 65.0 \\
R & 2,715.96 & 41,713 & 99,181 & 4.6 \\
C+P & 5,257.08 & 518,896 & 1,410,323 & 65.2 \\
C+R & 3,243.40 & 56,547 & 121,075 & 5.6 \\
P+R & 5,186.96 & 521,879 & 1,428,492 & 66.1 \\
C+P+R & 5,284.76 & 527,323 & 1,434,053 & 66.3 \\
\bottomrule
\end{tabular}
\endgroup
\end{table}

\begin{figure}[!htbp]
\centering
\includegraphics[width=\linewidth,height=0.76\textheight,keepaspectratio]{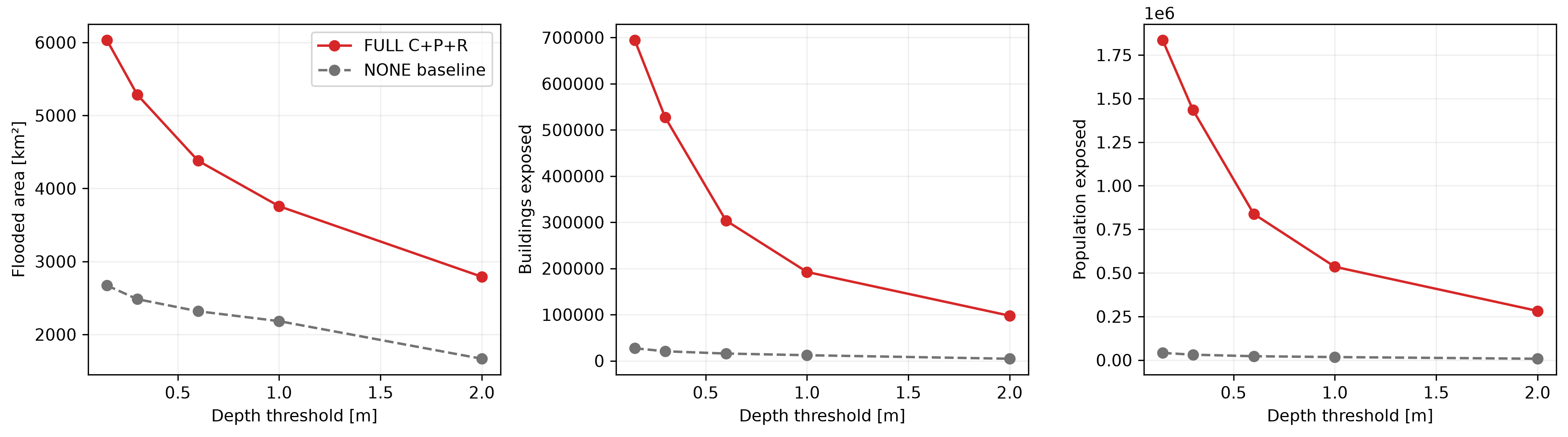}
\caption{Flooded area, exposed buildings, and exposed population for the full compound and background scenarios across the five reporting thresholds.}
\label{fig:7}
\end{figure}

Coastal and river forcing produce much smaller exposure totals when acting independently. Their importance nevertheless cannot be inferred from the single-driver simulations alone because interactions alter the combined response. The complete factorial experiment therefore provides the basis for the exact attribution in Section~\ref{sec:4.4.2} and the interaction analysis in Section~\ref{sec:4.4.3}.

The numerical decomposition is internally consistent. The exact Shapley reconstruction closes with an RMSE of \(9.17 \times 10^{- 8}\) m and a maximum absolute error of \(1.91 \times 10^{- 6}\) m. The factorial interaction decomposition has an RMSE of \(2.55 \times 10^{- 7}\) m, all far below the flood-depth thresholds used for impact analysis.

\subsubsection{Exact driver attribution}
\label{sec:4.4.2}

\begin{figure}[!htbp]
\centering
\includegraphics[width=\linewidth,height=0.8\textheight,keepaspectratio]{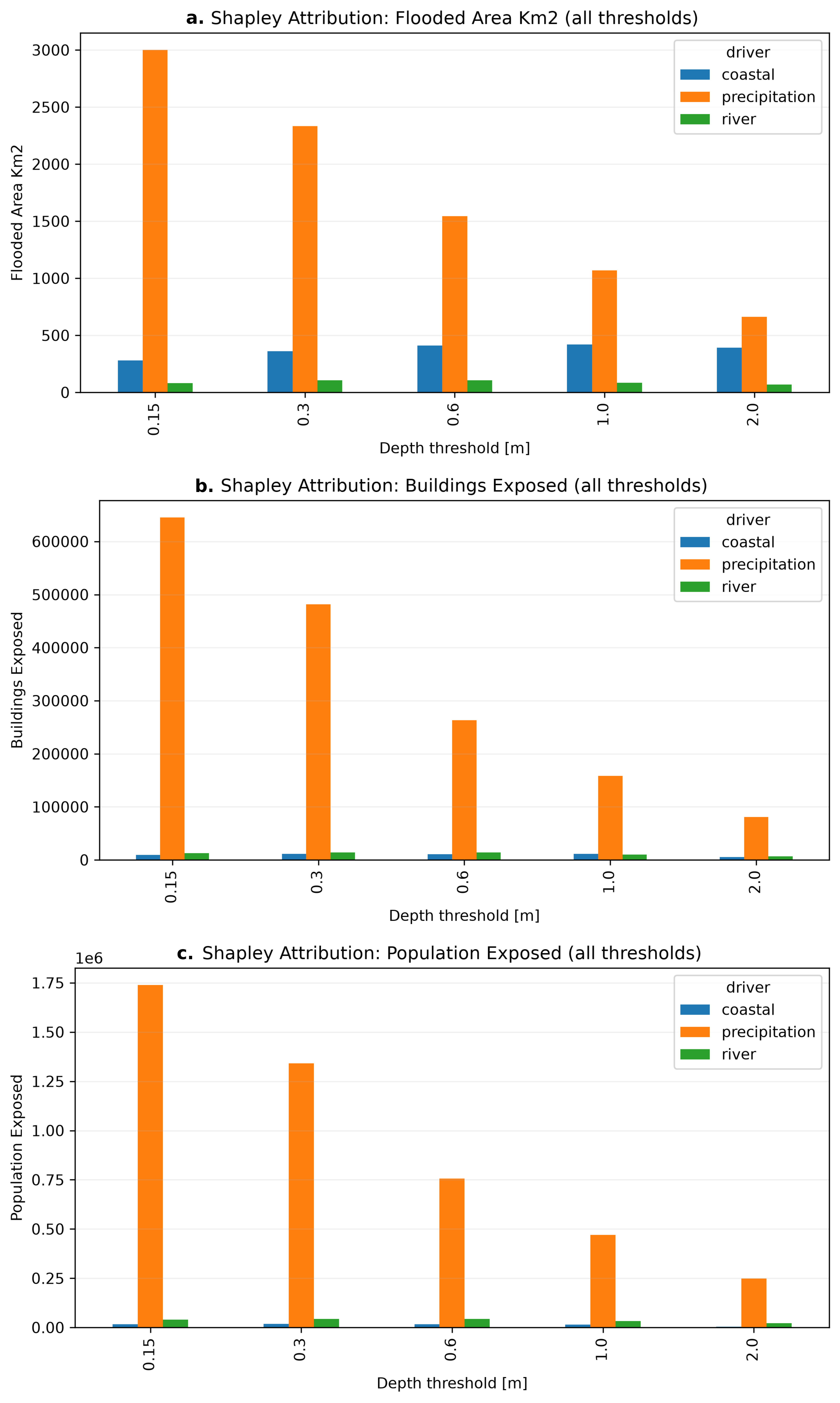}
\caption{Exact Shapley attribution across flood-depth thresholds. The panels compare driver contributions to flooded area, exposed buildings, and exposed population.}
\label{fig:8}
\end{figure}

Exact Shapley attribution shows that precipitation is the largest contributor to event-induced flooded area, building exposure, and population exposure across all five depth thresholds (Table~\ref{tab:8} and Figure~\ref{fig:8}). At 0.15 m, precipitation contributes 89.3\% of event-induced flooded area, 96.7\% of exposed buildings, and 96.9\% of exposed population.

The attribution changes with flood depth. At 2.0 m, precipitation still contributes 59.0\% of flooded area, 86.8\% of building exposure, and 90.5\% of population exposure. In contrast, the coastal contribution to flooded area increases from 8.3\% at 0.15 m to 34.9\% at 2.0 m. The river contribution remains smaller but increases to 6.1\% of flooded area and 8.1\% of population exposure at the deepest threshold.

This divergence between flooded area and societal exposure becomes increasingly important with depth. Coastal forcing contributes a substantial fraction of the deepest inundated area but a much smaller fraction of building and population exposure. This distinction is examined further using DIS in Section~\ref{sec:4.4.5}.

\begin{table}[!htbp]
\centering
\caption{Exact Shapley contributions of coastal, precipitation, and river forcing to event-induced flooded area, building exposure, and population exposure.}
\label{tab:8}
\medskip\par\textbf{\textbf{A. Flooded area}}\par\smallskip
\begingroup
\small
\setlength{\tabcolsep}{4.0pt}
\renewcommand{\arraystretch}{1.18}
\begin{tabular}{@{}>{\raggedright\arraybackslash}p{\dimexpr 0.13934\linewidth-3.344pt\relax}>{\raggedright\arraybackslash}p{\dimexpr 0.28689\linewidth-6.885pt\relax}>{\raggedright\arraybackslash}p{\dimexpr 0.28689\linewidth-6.885pt\relax}>{\raggedright\arraybackslash}p{\dimexpr 0.28689\linewidth-6.885pt\relax}@{}}
\toprule
\textbf{Depth (m)} & \textbf{Coastal (km\textsuperscript{2})} & \textbf{Precipitation (km\textsuperscript{2})} & \textbf{River (km\textsuperscript{2})} \\
\midrule
0.15 & 279.5 (8.3\%) & 3,000.1 (89.3\%) & 81.5 (2.4\%) \\
0.30 & 360.4 (12.9\%) & 2,333.8 (83.3\%) & 106.7 (3.8\%) \\
0.60 & 411.0 (19.9\%) & 1,543.4 (74.9\%) & 107.4 (5.2\%) \\
1.00 & 420.2 (26.7\%) & 1,068.8 (67.9\%) & 85.1 (5.4\%) \\
2.00 & 392.2 (34.9\%) & 663.2 (59.0\%) & 68.0 (6.1\%) \\
\bottomrule
\end{tabular}
\endgroup
\medskip\par\textbf{\textbf{B. Buildings exposed}}\par\smallskip
\begingroup
\small
\setlength{\tabcolsep}{4.0pt}
\renewcommand{\arraystretch}{1.18}
\begin{tabular}{@{}>{\raggedright\arraybackslash}p{\dimexpr 0.13934\linewidth-3.344pt\relax}>{\raggedright\arraybackslash}p{\dimexpr 0.28689\linewidth-6.885pt\relax}>{\raggedright\arraybackslash}p{\dimexpr 0.28689\linewidth-6.885pt\relax}>{\raggedright\arraybackslash}p{\dimexpr 0.28689\linewidth-6.885pt\relax}@{}}
\toprule
\textbf{Depth (m)} & \textbf{Coastal (buildings)} & \textbf{Precipitation (buildings)} & \textbf{River (buildings)} \\
\midrule
0.15 & 9,736 (1.5\%) & 645,078 (96.7\%) & 12,609 (1.9\%) \\
0.30 & 11,283 (2.2\%) & 481,317 (95.0\%) & 14,076 (2.8\%) \\
0.60 & 11,062 (3.8\%) & 262,996 (91.3\%) & 13,998 (4.9\%) \\
1.00 & 11,312 (6.3\%) & 158,638 (88.0\%) & 10,298 (5.7\%) \\
2.00 & 5,540 (5.9\%) & 80,853 (86.8\%) & 6,802 (7.3\%) \\
\bottomrule
\end{tabular}
\endgroup
\medskip\par\textbf{\textbf{C. Population exposed}}\par\smallskip
\begingroup
\small
\setlength{\tabcolsep}{4.0pt}
\renewcommand{\arraystretch}{1.18}
\begin{tabular}{@{}>{\raggedright\arraybackslash}p{\dimexpr 0.13934\linewidth-3.344pt\relax}>{\raggedright\arraybackslash}p{\dimexpr 0.28689\linewidth-6.885pt\relax}>{\raggedright\arraybackslash}p{\dimexpr 0.28689\linewidth-6.885pt\relax}>{\raggedright\arraybackslash}p{\dimexpr 0.28689\linewidth-6.885pt\relax}@{}}
\toprule
\textbf{Depth (m)} & \textbf{Coastal (people)} & \textbf{Precipitation (people)} & \textbf{River (people)} \\
\midrule
0.15 & 16,299 (0.9\%) & 1,738,255 (96.9\%) & 39,826 (2.2\%) \\
0.30 & 17,454 (1.2\%) & 1,341,472 (95.6\%) & 43,969 (3.1\%) \\
0.60 & 17,354 (2.1\%) & 755,626 (92.6\%) & 42,786 (5.2\%) \\
1.00 & 14,740 (2.8\%) & 470,743 (90.9\%) & 32,212 (6.2\%) \\
2.00 & 4,097 (1.5\%) & 248,219 (90.5\%) & 22,094 (8.1\%) \\
\bottomrule
\end{tabular}
\endgroup
\end{table}

\subsubsection{Nonlinear interactions and compound amplification}
\label{sec:4.4.3}

The factorial decomposition shows that the Harvey response is not additive. At the 0.30 m threshold, the coastal-precipitation interaction reduces flooded area by 568.8 km² relative to the sum of the corresponding main effects. The precipitation-river and coastal-river interactions are also negative, while the three-way \(C \times P \times R\) term contributes +139.2 km² (Table~\ref{tab:9}). Similar interaction signs occur for building and population exposure.

\begin{table}[!htbp]
\centering
\caption{Factorial main effects and interaction terms for flooded area, building exposure, and population exposure at the 0.30 m threshold.}
\label{tab:9}
\begingroup
\small
\setlength{\tabcolsep}{4.0pt}
\renewcommand{\arraystretch}{1.18}
\begin{tabular}{@{}>{\raggedright\arraybackslash}p{\dimexpr 0.36066\linewidth-8.656pt\relax}>{\raggedright\arraybackslash}p{\dimexpr 0.21311\linewidth-5.115pt\relax}>{\raggedright\arraybackslash}p{\dimexpr 0.21311\linewidth-5.115pt\relax}>{\raggedright\arraybackslash}p{\dimexpr 0.21311\linewidth-5.115pt\relax}@{}}
\toprule
\textbf{Factorial term} & \textbf{Flooded area (km\textsuperscript{2})} & \textbf{Buildings} & \textbf{Population} \\
\midrule
Main coastal & 669.5 (23.9\%) & 18,462 (3.6\%) & 33,163 (2.4\%) \\
Main precipitation & 2,672.6 (95.4\%) & 493,199 (97.3\%) & 1,373,782 (97.9\%) \\
Main river & 232.2 (8.3\%) & 21,066 (4.2\%) & 68,023 (4.8\%) \\
Coastal x precipitation & -568.8 (-20.3\%) & -13,412 (-2.6\%) & -27,779 (-2.0\%) \\
Coastal x river & -142.0 (-5.1\%) & -3,628 (-0.7\%) & -11,268 (-0.8\%) \\
Precipitation x river & -201.6 (-7.2\%) & -13,033 (-2.6\%) & -44,471 (-3.2\%) \\
Coastal x precipitation x river & 139.2 (5.0\%) & 4,022 (0.8\%) & 11,446 (0.8\%) \\
\bottomrule
\end{tabular}
\endgroup
\end{table}

These interaction terms describe departures from additivity. Compound amplification addresses a related but different question by comparing the full compound simulation with the largest single-driver response. The additional flooded area increases from 74.20 km² at the 0.15 m threshold to 384.64 km² at 2.0 m. When expressed as a percentage of the full CPR flooded area, amplification increases from 1.23\% to 13.78\% with depth (Table~\ref{tab:10}).

\begin{table}[!htbp]
\centering
\caption{Compound amplification calculated as the full CPR response minus the largest single-driver response across flood-depth thresholds. Values in parentheses express the amplification as a percentage of the full CPR value.}
\label{tab:10}
\begingroup
\small
\setlength{\tabcolsep}{4.0pt}
\renewcommand{\arraystretch}{1.18}
\begin{tabular}{@{}>{\raggedright\arraybackslash}p{\dimexpr 0.14286\linewidth-3.429pt\relax}>{\raggedright\arraybackslash}p{\dimexpr 0.28571\linewidth-6.857pt\relax}>{\raggedright\arraybackslash}p{\dimexpr 0.28571\linewidth-6.857pt\relax}>{\raggedright\arraybackslash}p{\dimexpr 0.28571\linewidth-6.857pt\relax}@{}}
\toprule
\textbf{Depth (m)} & \textbf{Added area (km\textsuperscript{2})} & \textbf{Added buildings} & \textbf{Added population} \\
\midrule
0.15 & 74.20 (1.23\%) & 9,804 (1.41\%) & 20,305 (1.11\%) \\
0.30 & 128.32 (2.43\%) & 13,477 (2.56\%) & 29,114 (2.03\%) \\
0.60 & 218.04 (4.98\%) & 15,774 (5.19\%) & 34,765 (4.15\%) \\
1.00 & 309.48 (8.24\%) & 13,420 (6.98\%) & 22,869 (4.27\%) \\
2.00 & 384.64 (13.78\%) & 8,208 (8.41\%) & 12,596 (4.46\%) \\
\bottomrule
\end{tabular}
\endgroup
\end{table}

\begin{figure}[!htbp]
\centering
\includegraphics[width=\linewidth,height=0.76\textheight,keepaspectratio]{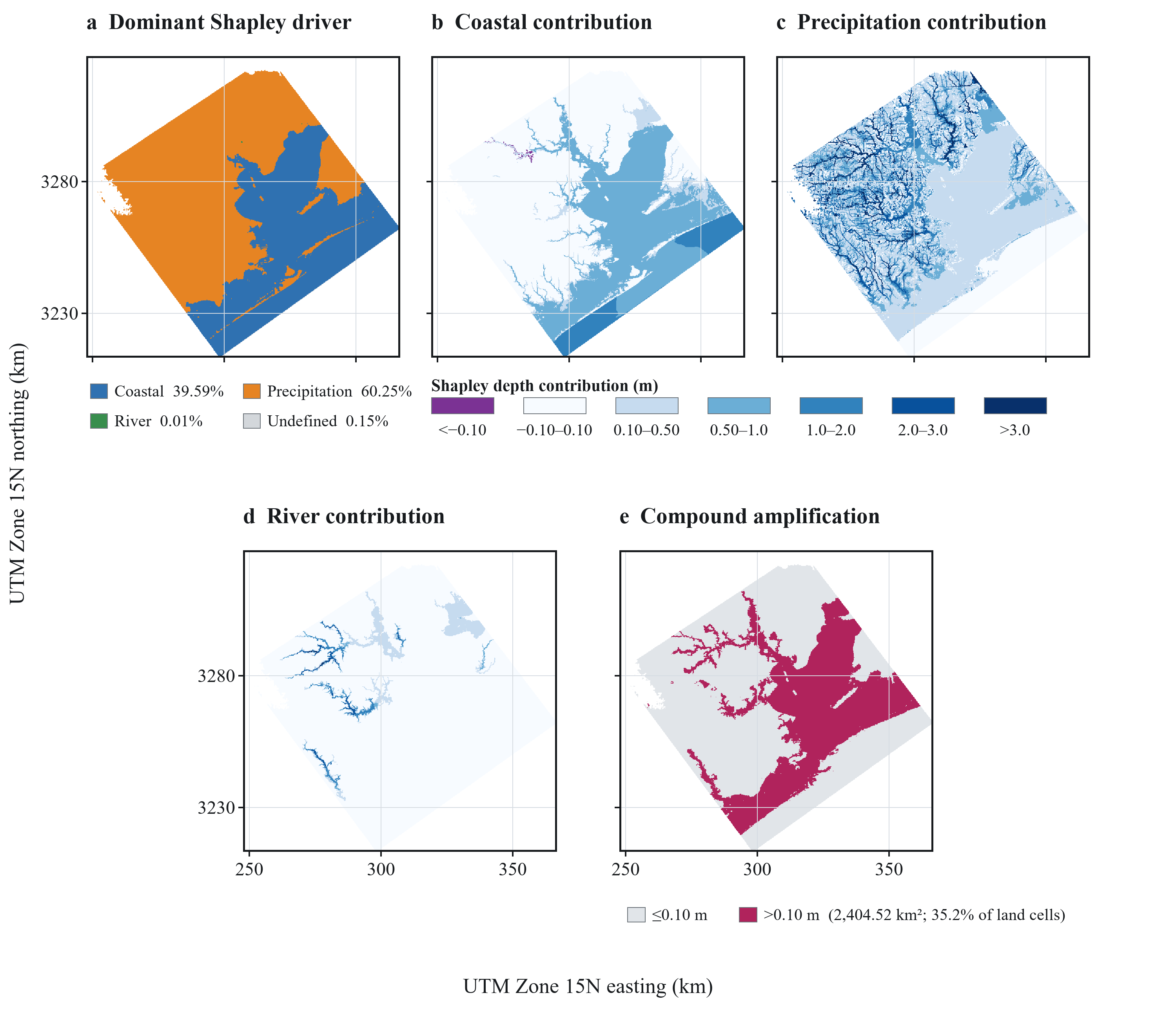}
\caption{Spatial Shapley attribution of event-maximum flood depth and compound amplification during Hurricane Harvey. (a) Dominant forcing at each land cell; precipitation and coastal forcing dominate 60.25\% and 39.59\% of the classified domain, respectively. (b--d) Shapley depth contributions (m) from coastal, precipitation, and river forcing, displayed using a common scale; negative values represent locally offsetting contributions. (e) Areas where compound amplification exceeds 0.10 m, covering 2,404.52 km², or 35.2\% of land cells. All panels use the same 200 m computational grid, land mask, map extent, and UTM Zone 15N coordinate system.}
\label{fig:9}
\end{figure}

Spatially, compound amplification greater than 0.10 m covers 2,404.52 km². Within these cells, the mean amplification is 0.235 m and the maximum reaches 1.994 m (Table~\ref{tab:11}). Figure~\ref{fig:9} connects this amplification pattern with the spatial Shapley decomposition. Precipitation dominates 60.25\% of classified land cells and coastal forcing 39.59\%, while river dominance is limited to a much smaller fraction of the domain.

The combination of Table~\ref{tab:8} to Table~\ref{tab:11}, and Figure~\ref{fig:9} therefore separates three different properties of the compound event: allocation of total hazard among drivers, nonlinear interaction among those drivers, and additional flooding produced by their combined occurrence.

\begin{table}[!htbp]
\centering
\caption{Spatial extent and magnitude of compound amplification. Relative amplification is defined as the amplification depth divided by the full CPR depth.}
\label{tab:11}
\begingroup
\footnotesize
\setlength{\tabcolsep}{2.0pt}
\renewcommand{\arraystretch}{1.18}
\begin{tabular}{@{}>{\raggedright\arraybackslash}p{\dimexpr 0.15000\linewidth-3.600pt\relax}>{\raggedright\arraybackslash}p{\dimexpr 0.10000\linewidth-2.400pt\relax}>{\raggedright\arraybackslash}p{\dimexpr 0.07000\linewidth-1.680pt\relax}>{\raggedright\arraybackslash}p{\dimexpr 0.17000\linewidth-4.080pt\relax}>{\raggedright\arraybackslash}p{\dimexpr 0.16000\linewidth-3.840pt\relax}>{\raggedright\arraybackslash}p{\dimexpr 0.16000\linewidth-3.840pt\relax}>{\raggedright\arraybackslash}p{\dimexpr 0.19000\linewidth-4.560pt\relax}@{}}
\toprule
\textbf{Criterion} & \textbf{Threshold} & \textbf{Units} & \textbf{Area (km\textsuperscript{2})} & \textbf{Mean amp. (m)} & \textbf{Max amp. (m)} & \textbf{Mean relative amp. (\%)} \\
\midrule
Absolute & 0.01 & m & 3,254.48 & 0.182 & 1.994 & 10.1 \\
Absolute & 0.05 & m & 2,572.92 & 0.224 & 1.994 & 11.8 \\
Absolute & 0.10 & m & 2,404.52 & 0.235 & 1.994 & 11.8 \\
Absolute & 0.25 & m & 524.16 & 0.454 & 1.994 & 19.2 \\
Relative & 0.05 & - & 1,843.88 & 0.240 & 1.994 & 15.4 \\
Relative & 0.10 & - & 1,045.36 & 0.281 & 1.994 & 21.6 \\
Relative & 0.25 & - & 265.40 & 0.355 & 1.994 & 40.7 \\
\bottomrule
\end{tabular}
\endgroup
\end{table}

\subsubsection{Flood persistence and duration attribution}
\label{sec:4.4.4}

The full compound event also produced long cumulative inundation durations. At the 0.30 m threshold, 4,866.28 km² accumulated at least 6 h above the threshold, and 4,184.92 km² accumulated at least 48 h above it. The corresponding number of buildings at locations exceeding the threshold for at least these cumulative durations decreased from 416,696 to 245,131. Even at the 1.0 m threshold, 3,294.56 km² and 116,420 buildings experienced at least 48 cumulative hours above the threshold (Table~\ref{tab:12}).

\begin{table}[!htbp]
\centering
\caption{Flooded area and building exposure accumulating at least 6, 12, 24, and 48 h above each flood-depth threshold in the full CPR scenario.}
\label{tab:12}
\medskip\par\textbf{\textbf{A. Persistent flooded area}}\par\smallskip
\begingroup
\small
\setlength{\tabcolsep}{4.0pt}
\renewcommand{\arraystretch}{1.18}
\begin{tabular}{@{}>{\raggedright\arraybackslash}p{\dimexpr 0.14050\linewidth-4.496pt\relax}>{\raggedright\arraybackslash}p{\dimexpr 0.21488\linewidth-6.876pt\relax}>{\raggedright\arraybackslash}p{\dimexpr 0.21488\linewidth-6.876pt\relax}>{\raggedright\arraybackslash}p{\dimexpr 0.21488\linewidth-6.876pt\relax}>{\raggedright\arraybackslash}p{\dimexpr 0.21488\linewidth-6.876pt\relax}@{}}
\toprule
\textbf{Depth (m)} & \textbf{\textgreater=6 h (km\textsuperscript{2})} & \textbf{\textgreater=12 h (km\textsuperscript{2})} & \textbf{\textgreater=24 h (km\textsuperscript{2})} & \textbf{\textgreater=48 h (km\textsuperscript{2})} \\
\midrule
0.15 & 5,568.20 & 5,330.40 & 5,063.04 & 4,661.44 \\
0.30 & 4,866.28 & 4,672.36 & 4,473.12 & 4,184.92 \\
0.60 & 4,150.12 & 4,038.24 & 3,916.40 & 3,719.08 \\
1.00 & 3,632.20 & 3,562.40 & 3,466.16 & 3,294.56 \\
2.00 & 2,697.12 & 2,648.48 & 2,566.64 & 2,421.60 \\
\bottomrule
\end{tabular}
\endgroup
\medskip\par\textbf{\textbf{B. Persistent buildings}}\par\smallskip
\begingroup
\small
\setlength{\tabcolsep}{4.0pt}
\renewcommand{\arraystretch}{1.18}
\begin{tabular}{@{}>{\raggedright\arraybackslash}p{\dimexpr 0.14050\linewidth-4.496pt\relax}>{\raggedright\arraybackslash}p{\dimexpr 0.21488\linewidth-6.876pt\relax}>{\raggedright\arraybackslash}p{\dimexpr 0.21488\linewidth-6.876pt\relax}>{\raggedright\arraybackslash}p{\dimexpr 0.21488\linewidth-6.876pt\relax}>{\raggedright\arraybackslash}p{\dimexpr 0.21488\linewidth-6.876pt\relax}@{}}
\toprule
\textbf{Depth (m)} & \textbf{\textgreater=6 h} & \textbf{\textgreater=12 h} & \textbf{\textgreater=24 h} & \textbf{\textgreater=48 h} \\
\midrule
0.15 & 594,853 & 526,256 & 459,174 & 369,923 \\
0.30 & 416,696 & 357,361 & 305,157 & 245,131 \\
0.60 & 240,972 & 212,884 & 188,568 & 160,428 \\
1.00 & 168,342 & 154,322 & 139,290 & 116,420 \\
2.00 & 84,247 & 75,872 & 65,444 & 50,182 \\
\bottomrule
\end{tabular}
\endgroup
\end{table}

Driver contributions to cumulative inundation duration differ from their contributions to maximum flood extent (Table~\ref{tab:13}). Precipitation contributes 80.1\% of event-induced flooded area-hours at 0.15 m, but this share decreases with depth. At 2.0 m, coastal forcing becomes the largest contributor to flooded area-hours, accounting for 56.0\%, compared with 37.5\% for precipitation.

Building-hours show a different pattern. Precipitation remains dominant across all thresholds, contributing 94.7\% at 0.15 m and 77.1\% at 2.0 m. Thus, the increasing coastal influence identified for deep flooded area in Section~\ref{sec:4.4.2} is even stronger for deep flood duration, while long-duration building exposure remains concentrated in precipitation-dominated areas.

\begin{table}[!htbp]
\centering
\caption{Shapley contributions to event-induced flooded area-hours and building-hours.}
\label{tab:13}
\begingroup
\small
\setlength{\tabcolsep}{4.0pt}
\renewcommand{\arraystretch}{1.18}
\begin{tabular}{@{}>{\raggedright\arraybackslash}p{\dimexpr 0.29032\linewidth-9.290pt\relax}>{\raggedright\arraybackslash}p{\dimexpr 0.13710\linewidth-4.387pt\relax}>{\raggedright\arraybackslash}p{\dimexpr 0.17742\linewidth-5.677pt\relax}>{\raggedright\arraybackslash}p{\dimexpr 0.21774\linewidth-6.968pt\relax}>{\raggedright\arraybackslash}p{\dimexpr 0.17742\linewidth-5.677pt\relax}@{}}
\toprule
\textbf{Duration metric} & \textbf{Depth (m)} & \textbf{Coastal (\%)} & \textbf{Precipitation (\%)} & \textbf{River (\%)} \\
\midrule
Flooded area-hours & 0.15 & 16.0 & 80.1 & 3.8 \\
Flooded area-hours & 0.30 & 27.8 & 66.6 & 5.6 \\
Flooded area-hours & 0.60 & 30.7 & 61.6 & 7.7 \\
Flooded area-hours & 1.00 & 35.1 & 56.3 & 8.6 \\
Flooded area-hours & 2.00 & 56.0 & 37.5 & 6.6 \\
Building-hours & 0.15 & 2.5 & 94.7 & 2.7 \\
Building-hours & 0.30 & 6.9 & 89.0 & 4.1 \\
Building-hours & 0.60 & 8.6 & 85.7 & 5.7 \\
Building-hours & 1.00 & 10.8 & 82.2 & 7.0 \\
Building-hours & 2.00 & 11.0 & 77.1 & 11.9 \\
\bottomrule
\end{tabular}
\endgroup
\end{table}

\subsubsection{Driver Impact Shift and social vulnerability}
\label{sec:4.4.5}

The difference between hazard footprint and consequence is quantified directly using DIS. Precipitation has positive DIS for both buildings and population at every depth threshold, indicating that its share of societal exposure is larger than its share of flooded area (Table~\ref{tab:14}). Population DIS increases from +7.61 percentage points at 0.15 m to +31.42 points at 2.0 m. Building DIS similarly increases from +7.39 to +27.72 points.

\begin{table}[!htbp]
\centering
\caption{DIS for exposed buildings and population relative to flooded area, and for building-hours relative to flooded area-hours. Positive values indicate that a driver has a larger share of the consequence metric than of its corresponding area metric; negative values indicate the opposite.}
\label{tab:14}
\begingroup
\small
\setlength{\tabcolsep}{4.0pt}
\renewcommand{\arraystretch}{1.18}
\begin{tabular}{@{}>{\raggedright\arraybackslash}p{\dimexpr 0.26613\linewidth-8.516pt\relax}>{\raggedright\arraybackslash}p{\dimexpr 0.13710\linewidth-4.387pt\relax}>{\raggedright\arraybackslash}p{\dimexpr 0.18548\linewidth-5.935pt\relax}>{\raggedright\arraybackslash}p{\dimexpr 0.22581\linewidth-7.226pt\relax}>{\raggedright\arraybackslash}p{\dimexpr 0.18548\linewidth-5.935pt\relax}@{}}
\toprule
\textbf{Impact metric} & \textbf{Depth (m)} & \textbf{Coastal} & \textbf{Precipitation} & \textbf{River} \\
\midrule
Buildings & 0.15 & -6.86 & 7.39 & -0.53 \\
Buildings & 0.30 & -10.64 & 11.67 & -1.03 \\
Buildings & 0.60 & -16.09 & 16.45 & -0.35 \\
Buildings & 1.00 & -20.42 & 20.11 & 0.31 \\
Buildings & 2.00 & -28.97 & 27.72 & 1.25 \\
Population & 0.15 & -7.41 & 7.61 & -0.20 \\
Population & 0.30 & -11.62 & 12.30 & -0.68 \\
Population & 0.60 & -17.81 & 17.77 & 0.03 \\
Population & 1.00 & -23.85 & 23.03 & 0.82 \\
Population & 2.00 & -33.42 & 31.42 & 2.00 \\
Building-hours & 0.15 & -13.51 & 14.60 & -1.09 \\
Building-hours & 0.30 & -20.93 & 22.35 & -1.42 \\
Building-hours & 0.60 & -22.12 & 24.07 & -1.96 \\
Building-hours & 1.00 & -24.29 & 25.87 & -1.59 \\
Building-hours & 2.00 & -44.92 & 39.61 & 5.31 \\
\bottomrule
\end{tabular}
\endgroup
\end{table}

Coastal forcing shows the opposite behavior. Its population DIS decreases from -7.41 percentage points at 0.15 m to -33.42 points at 2.0 m, showing that its growing contribution to deep flooded area does not translate proportionally into population exposure. River forcing remains comparatively close to proportional at shallow thresholds and becomes moderately consequence-heavy at greater depths.

These results connect directly with Sections~\ref{sec:4.4.2} and \ref{sec:4.4.4}. Coastal forcing becomes increasingly important for deep and persistent inundation, whereas precipitation remains more strongly concentrated in developed and populated portions of the domain. DIS expresses this difference explicitly rather than inferring it from separate hazard and exposure tables.

\begin{figure}[!htbp]
\centering
\includegraphics[width=\linewidth,height=0.8\textheight,keepaspectratio]{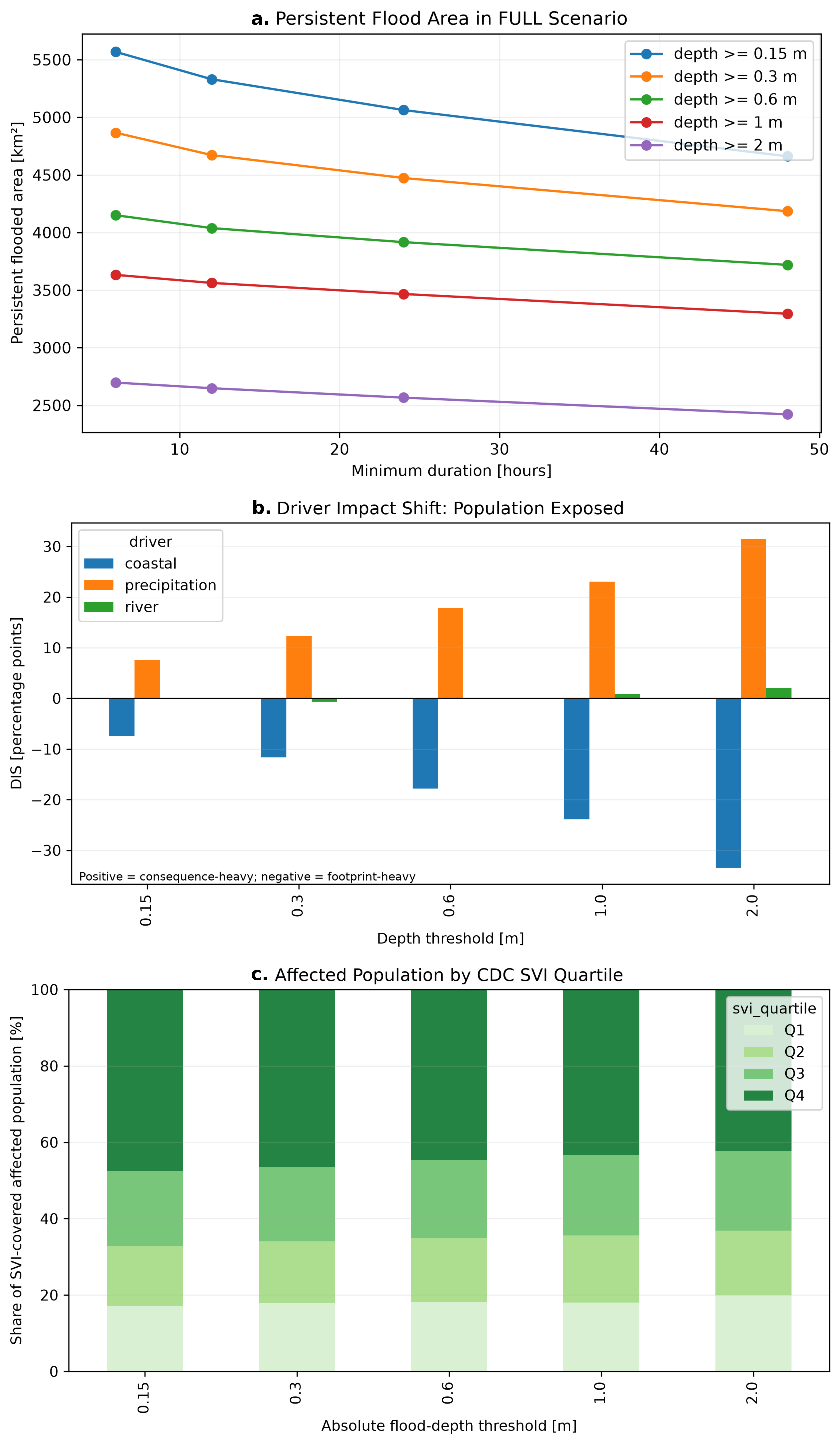}
\caption{Flood duration and consequence distribution. Panel (a) shows flooded area meeting the cumulative inundation-duration criteria; panel (b) shows population DIS; and panel (c) shows the distribution of SVI-covered affected population among CDC/ATSDR SVI quartiles across the common flood-depth thresholds.}
\label{fig:10}
\end{figure}

The social-vulnerability analysis provides an additional description of where the simulated impacts occur. CDC/ATSDR SVI coverage is at least 99.3\% of the affected population across all thresholds (Table~\ref{tab:15} and Figure~\ref{fig:10}). The highest-vulnerability quartile contains the largest absolute number of affected residents, in part because it contains 47.8\% of the baseline population in the analysis domain. Its burden concentration ratio remains below 1.0 and decreases from 0.995 at 0.15 m to 0.886 at 2.0 m. In contrast, Q1 and Q2 become relatively more concentrated at greater flood depths, reaching burden ratios of 1.169 and 1.105 at 2.0 m.

\begin{table}[!htbp]
\centering
\caption{CDC/ATSDR SVI burden concentration ratios across flood-depth thresholds.}
\label{tab:15}
\begingroup
\footnotesize
\setlength{\tabcolsep}{2.0pt}
\renewcommand{\arraystretch}{1.18}
\begin{tabular}{@{}>{\raggedright\arraybackslash}p{\dimexpr 0.13000\linewidth-3.120pt\relax}>{\raggedright\arraybackslash}p{\dimexpr 0.23000\linewidth-5.520pt\relax}>{\raggedright\arraybackslash}p{\dimexpr 0.10000\linewidth-2.400pt\relax}>{\raggedright\arraybackslash}p{\dimexpr 0.10000\linewidth-2.400pt\relax}>{\raggedright\arraybackslash}p{\dimexpr 0.10000\linewidth-2.400pt\relax}>{\raggedright\arraybackslash}p{\dimexpr 0.10000\linewidth-2.400pt\relax}>{\raggedright\arraybackslash}p{\dimexpr 0.24000\linewidth-5.760pt\relax}@{}}
\toprule
\textbf{Depth (m)} & \textbf{Affected population} & \textbf{Q1} & \textbf{Q2} & \textbf{Q3} & \textbf{Q4} & \textbf{SVI coverage (\%)} \\
\midrule
0.15 & 1,835,688 & 0.999 & 1.029 & 0.990 & 0.995 & 99.7 \\
0.30 & 1,434,053 & 1.046 & 1.060 & 0.982 & 0.972 & 99.6 \\
0.60 & 838,162 & 1.061 & 1.103 & 1.024 & 0.936 & 99.4 \\
1.00 & 535,215 & 1.054 & 1.151 & 1.058 & 0.908 & 99.3 \\
2.00 & 282,241 & 1.169 & 1.105 & 1.048 & 0.886 & 99.6 \\
\bottomrule
\end{tabular}
\endgroup
\end{table}

All in all, the attribution results show that the physical importance of a flood driver depends on the quantity being examined. Precipitation dominates regional exposure, coastal forcing becomes increasingly important for deep and persistent inundation, and nonlinear interactions increase the CCF footprint beyond that produced by any individual driver. The DIS results further show that a driver\textquotesingle s contribution to flooded area does not necessarily represent its contribution to buildings or population. These distinctions provide the basis for the broader interpretation of AutoCF and its limitations in Section~\ref{sec:5}.

The Harvey analysis demonstrates AutoCF's fully operational impact attribution workflow. The reported attribution magnitudes are resolution dependent because the spatial resolution of the hydrodynamic grid and the terrain, forcing, building, and population datasets governs the representation of flow pathways and exposed assets. Therefore, finer-resolution inputs can change the numerical driver shares reported here while leaving the attribution framework unchanged.

\section{Discussion}
\label{sec:5}

\subsection{From automated model setup to end-to-end CCF analysis}
\label{sec:5.1}

Automation in flood modeling has increasingly reduced the effort required to prepare terrain, forcing, boundaries, and numerical models. Rapid model-setup tools and globally applicable frameworks have made it possible to construct flood models more consistently across regions \citep{eilander2023a,sosa2020,vanormondt2020}. AutoCF advances this direction by extending automation beyond model construction and simulation to include observational evaluation, exposure analysis, CCF attribution, and interpretation within a unified workflow.

This integration is important because the scientific analysis does not end when a flood simulation is completed. Validation requires consistent treatment of observations and vertical datums, while impact analysis requires the simulated hazard to be connected with buildings, population, roads, and flood duration. Driver attribution adds another layer by requiring multiple simulations to be constructed and compared under a common domain, configuration, and analysis basis. AutoCF maintains these connections through a single project structure, so the transition from environmental inputs to modeled flooding and then to impacts remains reproducible and traceable.

The attribution component is particularly important in extending the workflow beyond conventional inundation analysis. AutoCF combines exact Shapley attribution with factorial interaction terms and compound amplification to distinguish individual driver contributions from nonlinear compound effects. These calculations are then propagated from physical flood depth to affected area, buildings, population, and duration. DIS adds a further layer by testing whether a driver\textquotesingle s contribution to societal consequences is proportional to its contribution to the flooded footprint.

The LLM-assisted component serves a different role. It provides access to workflow guidance, documentation, and saved scientific results without performing the numerical analyses itself. This keeps the interaction layer separate from deterministic calculations while making a complex modeling and attribution workflow easier to navigate.

\subsection{Flood-driver importance depends on the consequence being measured}
\label{sec:5.2}

The Hurricane Harvey experiment shows that the importance of a CCF driver depends strongly on the quantity used to measure it. Precipitation dominates regional exposure, but its relative contribution to flooded area decreases as the depth threshold increases. Its share of event-induced flooded area declines from 89.3\% at 0.15 m to 59.0\% at 2.0 m, while the coastal contribution increases from 8.3\% to 34.9\%. In contrast, precipitation still accounts for 86.8\% of exposed buildings and 90.5\% of exposed population at 2.0 m (Figure~\ref{fig:8}; Table~\ref{tab:8}). This divergence shows that the driver controlling the physical flood footprint is not necessarily the driver controlling societal exposure.

The contrast becomes stronger when flood persistence is considered. At 2.0 m, coastal forcing becomes the largest contributor to flooded area-hours, with a 56.0\% share, whereas precipitation still contributes 77.1\% of building-hours (Table~\ref{tab:13}; Figure \ref{fig:10}a). The driver controlling deep and persistent inundation can therefore differ from the driver controlling persistent exposure. This distinction matters because attribution based only on maximum flooded area can underrepresent how different drivers influence the duration and location of impacts.

The factorial results also show that the three drivers do not combine additively. At the 0.30 m threshold, the pairwise interactions are negative, while the three-way interaction is positive (Table~\ref{tab:9}). At the same time, compound amplification increases with depth: the additional flooded area relative to the strongest single-driver scenario rises from 1.23\% of the full-event footprint at 0.15 m to 13.78\% at 2.0 m (Table~\ref{tab:10}). Spatially, amplification greater than 0.10 m extends across 2,404.52 km² (Figure~\ref{fig:9}; Table~\ref{tab:11}). The increasing amplification at larger depths indicates that compound forcing becomes progressively more important for the severe part of the flood distribution, even though precipitation remains the largest contributor to overall exposure.

DIS makes the difference between hazard contribution and consequence contribution explicit. Precipitation has positive DIS for buildings and population at every tested depth, whereas coastal forcing has increasingly negative DIS (Table~\ref{tab:14}; Figure \ref{fig:10}b). For population, precipitation DIS increases from +7.61 percentage points at 0.15 m to +31.42 points at 2.0 m, while the coastal value decreases from \ensuremath{-}7.41 to \ensuremath{-}33.42 points. In Harvey, precipitation-driven flooding is therefore disproportionately concentrated in populated and built areas, whereas coastal forcing contributes more strongly to the physical footprint than to the corresponding societal exposure. This extends previous CCF attribution studies by quantifying how driver importance shifts when the analysis moves from hazard extent to consequences \citep{eilander2023b,grimley2025}.

These results extend recent CCF attribution research from decomposition of the physical flood response to attribution of its consequences. For example, \citep{chowdhury2026} used a Shapley value-based framework across all combinations of tide, storm surge, discharge, and precipitation to quantify their spatial and temporal contributions to compound water levels. Their analysis demonstrates how complete combinatorial attribution can resolve nonlinear driver contributions to the hydrodynamic response. AutoCF carries this framework substantially further, shifting the analysis from hazard attribution to consequence attribution by propagating driver contributions from flood depth to flooded area, buildings, population, flooded area-hours, and building-hours, thereby identifying which processes control not only the physical hazard but also the magnitude and persistence of societal consequences. DIS adds a further distinction by directly measuring whether a driver\textquotesingle s share of consequence is disproportionate to its share of the physical flood footprint. This consequence-centered attribution is important because, as demonstrated here, the dominant driver of CCF hazard can differ markedly from the dominant driver of exposure and persistent impact.

\subsection{Reliability and reproducibility of automated CCF modeling}
\label{sec:5.3}

An automated modeling workflow needs to demonstrate both observational credibility and numerical reproducibility. The Harvey benchmark provides evidence for both. Across eight NOAA stations, the automatically assembled simulation produces a median RMSE of 0.147 m and a median correlation of 0.951. The independent HWM comparison also shows strong spatial correspondence, with \(r = 0.942\) across 55 eligible observations. These results indicate that the automated selection and harmonization of terrain, forcing, boundaries, and observations can support a credible regional representation of a complex CCF event.

The cross-platform experiments provide a separate test of reproducibility. The domain-wide CPU-GPU maximum water-level comparison gives an MAE of 0.003 m, and 98.8\% of common cells agree within 0.01 m. Windows and macOS hydrographs also remain closely aligned at shared output times. This consistency is important for AutoCF because the workflow is intended to operate across personal computers, containerized environments, and HPC systems. Portability therefore needs to preserve the scientific solution, not only the ability to execute the software.

The largest localized differences occur near Manchester, which also shows the largest observational discrepancy. This provides a useful example of what automated evaluation can reveal. Most of the regional solutions are stable across platforms, while a limited area remains sensitive to local hydraulic representation. Such locations can be identified for targeted refinement or improved local data rather than requiring changes to the entire model.

Computational efficiency also matters because exact three-driver attribution requires repeated simulations. The HPC/GPU realization reduced the Harvey runtime from 50.28 min on Windows to 9.52 min. However, 89.1\% of the HPC elapsed time was associated with output writing rather than the momentum and continuity calculations. For repeated attribution experiments, efficient output handling and parallel scenario execution may therefore provide as much benefit as further solver acceleration.

\subsection{Geographic applicability, data dependence, and current limitations}
\label{sec:5.4}

AutoCF was designed around a combination of globally available datasets and replaceable user-defined inputs. Built-in global pathways cover terrain, atmospheric forcing, coastal water levels, river discharge, land cover, population, and event information, while local datasets can replace these sources when higher-resolution or better-validated information is available. This hybrid structure is important because the availability of globally consistent data has improved substantially, but detailed coastal bathymetry, river geometry, discharge observations, and datum-controlled water levels remain unevenly distributed.

The Harvey application provides a data-rich benchmark in which most components of AutoCF can be evaluated simultaneously. The U.S. setting provides dense hydrologic and oceanographic observations, HWMs, detailed terrain and soil information, building footprints, population data, and social-vulnerability information. Future applications in data-sparse regions will place greater emphasis on combining AutoCF\textquotesingle s global datasets with locally acquired observations and terrain information. The ability to substitute user-defined inputs is therefore central to applying the same workflow across regions with different data availability.

Several limitations remain in the current implementation. The attribution framework groups ocean water level, wind, and atmospheric pressure into a single coastal driver, so their individual effects are not separated. Exact attribution for three drivers requires seven additional simulations beyond the full event, and the computational requirement will increase rapidly if more driver groups are introduced. Exposure estimates also inherit uncertainty from population and building datasets, while the current SVI analysis is specific to U.S. applications. Historical simulations can additionally be affected when present-day terrain products differ from conditions during the event, and the current production evaluation does not yet include satellite-based inundation extent. These limitations identify clear areas for extending the framework without changing its underlying workflow.

Although the present implementation is demonstrated for historical event hindcasting, the workflow is organized around replaceable forcing inputs and is therefore adaptable to other application modes. For real-time or short-range forecasting, historical precipitation, coastal water level, and river discharge inputs could be replaced by corresponding forecast products, while retaining the same model construction, simulation, inundation, exposure, and post-processing workflow. Similarly, hazard and risk analyses could use synthetic or probabilistic combinations of coastal water levels, precipitation, and river discharge to generate ensembles of CCF scenarios and associated impacts. These extensions would require dedicated interfaces for forecast or synthetic forcing products and separate validation, but would not require changing the underlying AutoCF workflow architecture.

The present AutoCF release provides the complete end-to-end workflow demonstrated in this study. Future development will expand the range of physical processes and spatial configurations that AutoCF can construct automatically. Support for SFINCS quadtree grids will enable locally refined cells along shorelines, narrow channels, levees, and densely developed areas while retaining coarser resolution elsewhere, improving the representation of critical flow pathways without the computational cost of a uniformly fine grid. Integration of the stationary SnapWave solver will add incident- and infragravity-wave energy propagation, wave breaking, and wave-induced setup, extending AutoCF to exposed coasts where waves contribute materially to total water levels and inundation. Automated representation of thin dams and weirs, culverts, pumps, and check valves will capture levees, barriers, drainage connections, and backflow controls, while Horton infiltration and subgrid storage-volume options will strengthen urban rainfall-runoff and green-infrastructure applications. These capabilities will build directly on the fully functional current system and allow future releases to tailor model physics, resolution, and infrastructure representation to regional, urban, and wave-exposed coastal applications.

\section{Conclusions}
\label{sec:6}

This study introduces AutoCF, an automated LLM-assisted ecosystem for CCF simulation, evaluation, and impact attribution. AutoCF does not replace the underlying hydrodynamic model. Instead, it integrates data acquisition and harmonization, model construction, execution, observational evaluation, impact analysis, driver attribution, and interpretation within a common reproducible workflow. The current implementation uses HydroMT-SFINCS for model construction and SFINCS for hydrodynamic simulation, while AutoCF provides the higher-level orchestration and analysis framework.

Hurricane Harvey was used as an integrated benchmark because it combines strong precipitation, river, and coastal forcing with extensive observations. The automatically assembled simulation reproduces the principal water-level response across the Houston-Galveston domain and captures the spatial variation of observed HWMs. Cross-platform tests further show close numerical agreement among Windows, macOS, and HPC implementations, while the GPU backend substantially reduces runtime for the repeated simulations required by attribution.

The factorial analysis shows that Harvey\textquotesingle s impacts cannot be summarized by flooded area alone. Precipitation dominates regional building and population exposure, while the coastal contribution increases for deeper and longer-lasting flooding. Nonlinear interactions further enlarge the CCF footprint. The introduced DIS metric quantifies an additional distinction between hazard and consequence by showing whether each driver\textquotesingle s impact share is larger or smaller than its flooded-area share.

The Harvey benchmark demonstrates the complete AutoCF workflow for historical CCF hindcasting in a data-rich U.S. setting. Global applications remain dependent on the quality of available terrain, forcing, observations, and exposure data, and AutoCF is therefore designed to combine globally available products with user-supplied local datasets. The same modular architecture also provides a basis for future extensions to real-time forecasting using meteorological, coastal, and river forecasts, and to hazard and risk analysis using synthetic or probabilistic forcing scenarios. The current release therefore provides a reproducible framework for event-based CCF reconstruction, evaluation, and driver-specific impact attribution, while retaining the flexibility needed for broader predictive and risk-based applications.

\section*{Author Contributions}

\textbf{Conceptualization}: S.R. and F.M.; \textbf{Methodology}: S.R. and F.M.; \textbf{Software}: S.R. and F.M.; \textbf{Validation}: S.R., F.M., N.L. and H.M.; \textbf{Formal Analysis}: S.R. and F.M.; \textbf{Investigation}: S.R. and F.M.; \textbf{Data Curation}: S.R. and F.M.; \textbf{Writing - Original Draft:} S.R.; \textbf{Visualization:} S.R. and F.M.; \textbf{Writing - Review and Editing:} N.L., H.M. and F.M.; \textbf{Supervision:} N.L.

\section*{Competing Interests}

The authors declare no conflict of interest.

\section*{Code and Data Availability}

The AutoCF software and accompanying documentation are available through the project website at \url{https://www.autocf.net}. The external datasets used for model construction, observational evaluation, and exposure analysis are described and cited in this manuscript and can be obtained from their respective providers.

\FloatBarrier
\begingroup
\raggedright

\endgroup
\end{document}